\documentclass[a4paper,12pt]{article}
\usepackage{graphicx,graphics,xcolor}
\usepackage{epsfig}
\usepackage{amsmath}
\usepackage{amssymb}
\usepackage{mathtools}
\usepackage[top=1cm, bottom=1.5cm, left=2cm, right=2cm]{geometry}
\usepackage{microtype}
\usepackage{lmodern}
\usepackage[utf8]{inputenc}
\usepackage{float}
\newcommand{\ds}{\displaystyle }

\newcommand{\beq}{\begin{equation} }
\newcommand{\eeq}{\end{equation}}

\begin{document}

\title{Four compartment epidemic model with retarded transition rates}
\author{ T\'eo Granger$^1$, Thomas M. Michelitsch$^1$\footnote{Corresponding, \, E-mail:\, thomas.michelitsch@sorbonne-universite.fr}, Michael Bestehorn$^2$, \\ Alejandro P. Riascos$^3$, Bernard A. Collet$^1$\\
$^1$ Sorbonne Universit\'e \\ Institut Jean le Rond d'Alembert 
CNRS UMR 7190 \\ 4 place Jussieu, 75252 Paris cedex 05, France
\\ 
$^2$
Brandenburgische Technische Universit\"at Cottbus-Senftenberg \\
Institut f\"ur Physik \\ Erich-Weinert-Stra{\ss}e 1, 03046 Cottbus, Germany
\\ 
$^3$
Instituto de F\'isica, Universidad Nacional Aut\'onoma de M\'exico\\
Apartado Postal 20-364, 01000 Ciudad de M\'exico, M\'exico
}
\maketitle
\begin{abstract}
{\small 
We study an epidemic model for a constant population by taking into account
four compartments of the individuals characterizing their states of health. Each individual is in one of the compartments susceptible (S); incubated -- infected yet not infectious (C), infected and infectious (I), and recovered -- immune (R). An infection is `visible' only when an individual is in state I. Upon infection, an individual performs the transition pathway
$\rm S \to C \to I \to R \to S$ remaining in each compartments C, I, and R
for certain random waiting time $t_C,t_I,t_R$, respectively. The waiting times for each compartment are independent and drawn from specific probability density functions (PDFs) introducing memory into the model. The first part of the paper is devoted to the macroscopic SCIRS model.
We derive memory evolution equations involving convolutions (time derivatives of general fractional type). We consider several cases. The memoryless case is represented by exponentially distributed waiting times. Cases of long waiting times with fat-tailed waiting time distributions are considered as well where the SCIRS evolution equations take the form of time-fractional ODEs.
We obtain formulae for the endemic equilibrium and a condition of its existence for cases when the waiting time PDFs have existing means. We analyze the stability of healthy and endemic equilibria and derive conditions for which the endemic state becomes oscillatory (Hopf) unstable.
In the second part, we implement a simple multiple random walker's approach (microscopic model of Brownian motion of $Z$ independent walkers)
with  random SCIRS waiting times into computer simulations.
Infections occur with a certain probability by collisions of walkers in compartments I and S.
We compare the endemic states predicted in the macroscopic model with the numerical results of the simulations and find accordance of high accuracy.
We conclude that a simple random walker's approach
offers an appropriate microscopic description for the macroscopic model.
SCIRS type models open a wide field of applications allowing the identification
of pertinent parameters governing the phenomenology of epidemic dynamics such as extinction, convergence to a stable endemic equilibrium, or persistent oscillatory behavior.}
\end{abstract}

\newpage
%%%%%%%%%%%%%%%%%%%%%%%%%%%%%%%%%%%%%%%%%%%%%%%%%%%%%%%%%%%%%%%%%%%%%%%%%%%%%%%%%%%%%%%%%%%%%%%%%
%\pagestyle{empty}
%
%
%\tableofcontents
%
%

\section{Introduction}
The origin of modern epidemic modeling started with the seminal work of Kermack and Mc Kendrick
almost a century ago \cite{kermack}. They introduced the so called
`SIR model' where SIR stands for three compartments characterizing the states of health of an individual, namely S $=$ susceptible, I $=$ infected, R $=$ recovered. The classical SIR model and various generalizations are able to capture some of the features of epidemic spreading of infectious diseases as observed in influenza, measles, mumps, and rubella.
In the meantime a huge burst of compartment models and generalizations of the SIR model have been introduced \cite{anderson,mart}.

While the interest in mathematical modeling of epidemic spreading
was growing continuously, it is unsurprising that the emergence of the COVID pandemics has launched an enhanced interest and urgent need in advanced epidemic modeling \cite{BelikGeiselBrockmann2011}.
Many models consider a set of nonlinear ordinary differential equations (ODEs) for the time evolution of the compartment populations where
a new direction is about to emerge by combining these models with approaches inspired by network science
\cite{Barabasi2016,Satorras-Vespigniany2001,Feng-et-al2020,RiascosMateos2021,MiRia-ISTE2019,BesRiascosMichel2020,besmi}
and fractional dynamics \cite{MetzlerKlafter2000,SandevChechkinMetzler2021,MainardiGorenfloScalas2004,fractalfract2020,MiRia-FCAA2020}.

It has turned out that in many cases epidemics including COVID exhibit quasi-periodic patterns and spontaneous new outbreaks even after longer times of inactivity. Persistent oscillatory characteristics in epidemic dynamics was already pointed out a long time ago in the work of Soper \cite{soper}.
Indeed, a major drawback of classical SIR type models without memory effects lies in their
incapacity to capture (persistent) oscillatory behavior.

The present paper aims to tackle this issue and
is a generalization of recent works \cite{BesRiascosMichel2020,besmi}. In \cite{besmi} is introduced a SIRS compartment model which takes into account a random duration of immunity R allowing a delayed transition R $\to$ S. What is found is that this model is able to capture persistent oscillatory behavior in the number of infected individuals. This feature comes along as oscillatory Hopf instability of the endemic state.
In the first part of our paper we extend this model to four compartments as relevant states of health in order to capture a larger variety of epidemics. We consider the four compartments  S -- susceptible;
 C -- infected but not yet infectious (in the incubation phase); I-- infected and infectious (ill); R-- recovered and immune. We assume random sojourn times (waiting times) $t_C,t_I,t_R$ an individual once infected spends in each of the compartments C, I, R. The sojourn time $t_C$
in compartment C is interpreted as incubation time, i.e. the delay between infection and the outbreak of the disease. The sojourn times $t_I$ is the duration of the disease (infected and infectious state)
and $t_R$ indicates the immunity period after recovery. It appears natural to assume that these variables are not fixed constants, but individually fluctuating
random variables drawn by specific distributions. With these assumptions we derive
the SCIRS evolution equations where we focus on several pertinent situations of waiting times with existing and non-existing mean. For waiting time probability density functions (PDFs) with existing mean we derive explicit formulae for the endemic equilibrium and
analyze its stability.

In the second part of the paper we combine our SCIRS model with a multiple random walker's approach which is implemented into computer simulations. We present a case study and give numerical evidence that a simple random walk approach (`Brownian motion') offers an appropriate description of the SCIRS dynamics with memory. 

Indeed, the uncertainty in the available data (for instance the number of infected individuals) in many real world epidemics is a major problem which strongly affects their predictability. Therefore, stochastic approaches which include hypotheses on randomness of the involved quantities, for instance on the allowed steps in a random walk of the individuals or in the model parameters (such as waiting times in the compartments or considering the transition rates as random variables)
offer powerful modeling tools dealing in a natural way with
the lack of available information. A stochastic approach with random transition rates has been presented in a recent work of Faranda et al. \cite{FarandaAlberti2022-SEIR} 
to model the second wave of COVID. In that model the epidemic dynamics has been described
in the framework of a stochastic SEIR (standing for Susceptible-Exposed-Infected-Recovered) compartment model with classical types of evolution equations where the transition rates are random variables drawn from a Gaussian distribution. In that work the compartment R accounts for the recovered (immune) and dead individuals where immunity is never ending (infinite immunity time). Hence there is no transition back to the susceptible state. The compartment E (exposed) corresponds to infected but not infectious individuals. As a consequence of the infinite immunity time the endemic state is stable and (in contrast to our SCIRS model) no persistent oscillations in the number of infections occur in their model. We discuss cases of infinite waiting times in the compartments in our model as a limiting case of infinite immunity duration (see Section \ref{endemic_gen}) where a stable endemic state with a constant population of individuals in compartment R in the limit of large times occurs, consistent with their result \cite{FarandaAlberti2022-SEIR}.

\section{Four compartment model}
\label{Four_compartment_model}
\subsection{Basic notions}
We consider a constant (time independent) total population of $Z=Z_S+Z_C+Z_I+Z_R$ individuals (`random walkers') where $Z_S(t),Z_C(t),Z_I(t),Z_R(t)$ indicate their numbers in the compartments S C I R at time $t$. Assuming a constant total population means that we focus here on the large class of epidemics for which the mortality as well as natural birth and death processes can be neglected, at least on relevant time scales of the epidemic dynamics for instance during occurrence of infection waves. More precisely, we assume that the waiting times $t_{C,I,R}$ are small
compared to the expected life times of the individuals.
We consider here first a continuous time model for $Z \gg 1$ walkers with the compartment population fractions $s(t)=\frac{Z_S(t)}{Z}$, $c(t)=\frac{Z_C(t)}{Z}$, $j(t)=\frac{Z_I(t)}{Z}$, $r(t)=\frac{Z_R(t)}{Z}$. 
We allow for each walker the following transition pathway:
\begin{center}
 {\large ${\rm S \to C \to I \to R \to S}$}
\end{center}
Intuitively we infer that allowing for persistent oscillatory and (quasi-) periodic behavior it is necessary to have a closed (cyclic) pathway S $\to \ldots \to $ S of transitions in a constant (time independent) total population as assumed in our model. However, this is not anymore true in models where the total population is allowed to vary such as considered in the article of Manfredi and Salinelli \cite{ManfrediSalinelli2002}. They demonstrated in this work that sustained oscillations may occur when being induced in an exponentially growing total population.
We assume that $t_C,t_I,t_R \geq 0$ are mutually {\it independent} random waiting times an individual spends (after infection) in the compartments C, I, R drawn from specific probability density functions (PDFs) which we prescribe by the kernels $K_C(t),K_I(t),K_R(t)$, respectively. 
These kernels have to be causal functions\footnote{A function $g(t)$ is called `causal' if $g(t) = 0$ for $t<0$, i.e. non-zero only for $t\geq 0$.} with (read $\mathbb{P}$ as `probability'), for instance
$$
\mathbb{P}(t_C \in [\tau,\tau+{\rm d}\tau]) =  K_C(\tau){\rm d}\tau , \hspace{1cm} t_C >0
$$
indicates the probability that the incubation time $t_C \in [\tau,\tau+{\rm d}\tau]$
which is clearly non-zero only for $\tau\geq 0$ as $t_C \geq 0$. 
In other words: Given a walker has entered compartment C at $\tau=0$, then
$K_C(\tau){\rm d}\tau$ indicates the probability that this walker leaves compartment C during the infinitesimal interval $[\tau,\tau+{\rm d}\tau]$ (by a transition C $\to$ I).
We introduce the infection rate ${\cal A}(t)={\cal A}(j(t),s(t),t) \geq 0$ (number of infections per walker and unit time or entry rate to the incubated compartment C at time $t$ having of units $sec^{-1}$). 
We point out that ${\cal A}(t)$ is not
a known given function of time $t$ but rather an implicit non-linear function of $s(t),j(t)$, namely
$${\cal A}(t) ={\cal A}[s(t),j(t)].$$
This quantity contains macroscopic information on the instantaneous number of collisions of individuals in the `reactive compartments' S and I and needs to be specified by model assumptions. It is important to point out that ${\cal A}(t)$ does not depend on the fractions $c(t),r(t)$ of the `invisible' compartments C and R.

We assume here the most simple nonlinear form ${\cal A}(t)= \beta j(t)s(t)$ where $\beta>0$ is a constant independent of time where $\beta^{-1}$ introduces a characteristic time scale. $\beta$ is also a macroscopic measure for the infection probability in a collision of an $S$ and an $I$ walker.
The microscopic picture which we will invoke subsequently and implement into computer simulations indeed is that the infection rate is driven by random
collisions between infected and susceptible independent random walkers. 

With these simple assumptions we can establish model equations governing the time evolution of the fractions $s(t),c(t),j(t),r(t)$. 
We consider the dynamics starting at time $t=0$ with an initial condition $s(0)= 1-j_0$, and $j(0)=j_0$ (many healthy and a few infected walkers) and no incubated or recovered walkers $c(0)=r(0)=0$.
Our (SCIRS)- model evolution equations have the following general structure
\begin{equation}
 \label{SCIRS-model}
 \begin{array}{clr}
\displaystyle  \frac{d}{dt}s(t) & = - {\cal A}(t)   +  ({\cal A} \star K_C \star K_I \star K_R)(t) & \\  \\
\displaystyle \frac{d}{dt}c(t) & = {\cal A}(t) - ({\cal A} \star K_C)(t) \\ \\
\displaystyle  \frac{d}{dt}j(t) & =  ({\cal A} \star K_C)(t) - ({\cal A} \star K_C \star K_I)(t) \\ \\
 \displaystyle  \frac{d}{dt}r(t) & = ({\cal A} \star K_C \star K_I)(t) -  
  ({\cal A} \star K_C \star K_I\star  K_R)(t)
 \end{array}
\end{equation}
The sum of these rates is vanishing due to $s(t)+c(t)+j(t)+r(t)=1$. 
In order that an epidemics starts, it is necessary that the globally healthy state becomes unstable when the initial state is close to the healthy state with $j_0 \sim\frac{1}{Z} \to 0+$, i.e. where for instance only one infected walker is present at $t=0$ (among a large population $Z$).
We analyze the stability of the healthy state subsequently in detail.
In (\ref{SCIRS-model}) we have employed the notation 
\beq
\label{causal}
(K \star f)(t) = \int_0^tK(t-\tau)f(\tau){\rm d}\tau , \hspace{1cm} t \geq 0
\eeq
for convolutions of causal functions $K(t),f(t)$.
Be aware that convolutions commute and are associative which can be seen by the representation ($t\geq 0$) of the multiple convolution 
\begin{equation}
\label{nfold-convol}
(K_1 \star K_2 \star \ldots \star K_n)(t)  =: 
\int_{-\infty}^{\infty}\ldots \int_{-\infty}^{\infty}
{\rm d}\tau_1\ldots {\rm d}\tau_n
\delta(t-\tau_1-\ldots -\tau_n) K_1(\tau_1)\ldots K_n(\tau_n)
\end{equation}
where $\delta(\ldots)$ indicates the Dirac $\delta$-function and (\ref{nfold-convol}) is non-zero only for $t\geq 0$ (causality of the $K_i(\tau)$).
Introducing the Laplace transform of our kernels which we define as
$${\hat K}(\lambda)=\int_0^{\infty} e^{-\lambda t}K(t){\rm d}t$$
where $\lambda$ indicates the Laplace variable. 
${\cal K}(\lambda)\big|_{\lambda=0}=1$ indicates that
the kernels are normalized PDFs. A further observation is worthy of mention.
Any (multiple) convolution of PDFs gives again a PDF. This can be easily seen
by integrating (\ref{nfold-convol}) over time to yield one. 
Therefore the kernels $K_C(t),\, (K_C\star K_I)(t),\, (K_C \star K_I\star  K_R)(t)$ which appear in (\ref{SCIRS-model}) all are (normalized and causal) PDFs representing, respectively the densities of the random variables $t_C$, $t_C+t_I$, and $t_C+t_I+t_R$.
We can see in (\ref{SCIRS-model}) that for each compartment where an individual remains for a random 
duration an additional convolution occurs describing the delayed transition out of the compartment. These delayed transitions
introduce memory into the rates of individuals leaving the compartments. 
We will give a careful account for this issue later (subsequent representation (\ref{SCIRS-model_memory})).

The interpretation of (\ref{SCIRS-model}) is as follows. In the first line ${\cal A}(t)$ is the rate of infections 
(transitions S $\to$ C) at time $t$. 
The second term $({\cal A} \star K_C \star K_I \star K_R)(t)$
is the rate of individuals losing their immunity (R $\to$ S) having undergone the full pathway of SCIRS transitions. 
In the second Eq. of (\ref{SCIRS-model}) ${\cal A}(t)$ re-appears as the entry rate to state C. The 
convolution $({\cal A} \star K_C)(t)$ is the rate of delayed transition C $\to$ I of individuals that fall ill at time $t$. This rate reappears in the third line of (\ref{SCIRS-model}) 
as entry rate into compartment I. Then $({\cal A} \star K_C \star K_I)(t)$ is the rate of recovery I $\to$ R reoccurring in third equation as entry rate into compartment R. Finally,  
$({\cal A} \star K_C \star K_I\star  K_R)(t)$ is the rate of individuals losing their immunity (transition R $\to$ S) and re-appears in the first equation as entry rate into S. 
Let us
consider for a moment the second equation in (\ref{SCIRS-model}) more closely 
\beq
\label{eqC}
\frac{d}{dt} c(t) = {\cal A}(t) - \int_0^tK_C(t-\tau){\cal A}(\tau){\rm d}\tau 
\eeq
For our convenience we introduce the `survival probability' $\Phi_{C,I,R}(t)$ for individuals in compartments C, I, R (also called `persistence probability' or `survival function'), e.g. \cite{MainardiGorenfloScalas2004} (and many others)
\begin{equation}
\label{survival_prob}
\Phi_{C,I,R}(t) = \mathbb{P}(t_{C,I,R} > t) = \int_t^{\infty}
K_{C,I,R}(\tau){\rm d}\tau = 1 -\int_0^t K_{C,I,R}(\tau){\rm d}\tau
\end{equation}
indicating the probability that an individual which is entering compartment C, I, R, respectively at $t'=0$
at time $t'=t$ still is (`survives') in this compartment.
(\ref{survival_prob}) is capturing all realizations with $t_{C,I,R} > t$ where $\frac{d}{dt}\Phi_{C,I,R}(t)=-K_{C,I,R}(t)$. We observe in this relation the initial condition
$\Phi_{C,I,R}(0)=1$ as a consequence of the normalization of the waiting time PDFs and is telling us that an individual is, with probability one, in a compartment at the instant when entering it. Further we have
$\Phi_{C,I,R}(t \to \infty) \to 0+$, i.e. an individual `survives' only a finite time in compartments C, I, R to enter eventually the susceptible state S.
We hence can integrate (\ref{eqC}) and rewrite as
\beq
\label{rewrite}
c(t) = \int_0^t\Phi_C(t-\tau){\cal A}(\tau){\rm d}\tau = (\Phi_C\star {\cal A})(t)
\eeq
where initial condition $c(0)=0$ is assumed. 
We can also verify this relation by its Laplace transform (subsequent second equation of (\ref{SCIRS-model-laplace})) and directly by
differentiating this expression with respect to $t$ and using the initial condition $\Phi_C(0)=1$.
We interpret (\ref{rewrite}) as follows: ${\cal A}(\tau){\rm d}\tau$ is the fraction of walkers entering C during $[\tau,\tau+{\rm d}\tau]$ and $\Phi_C(t-\tau)$ is the (survival-) probability that 
this fraction still is in C after a delay of $t_C=t-\tau$, i.e. at time instant $t$ where this expression sums up over the complete history of entries into C from $0$ to time $t$.
With these considerations we can rewrite our SCIRS Eqs. (\ref{SCIRS-model}) in the equivalent integral form 
\beq
 \label{SCIRS-model_integral}
 \begin{array}{clr}
\displaystyle  s(t) & = 1 - c(t) - j(t) - r(t)   & \\  \\
\displaystyle c(t) & =   (\Phi_C\star {\cal A})(t) \\ \\
\displaystyle  j(t) & = j_0+ ({\cal A} \star K_C \star \Phi_{I})(t)   \\ \\
 \displaystyle  r(t) & = ({\cal A} \star K_C \star K_I \star \Phi_{R})(t)
 \end{array} ,\hspace{1cm} (t\geq 0)
\eeq
where the initial conditions $s(0)=1-j_0$, $c(0)=0$, $j(0)=j_0$, $r(0)=0$ are assumed. 
\subsection{Memoryless case -- exponentially distributed waiting times}
\label{momoryless}
It is worthy to consider exponential waiting time kernels 
$K_{C,I,R}(t)=\xi_{C,I,R} e^{-t\xi_{C,I,R}}$ ($\xi_{C,I,R}^{-1} = \langle t_{C,I,R}\rangle$) where the survival functions are also exponentials $\Phi_{C,I,R}(t)= e^{-t\xi_{C,I,R}}$ thus
$K_{C,I,R}(t) = \xi_{C,I,R} \Phi_{C,I,R}(t)$. Substituting 
this into (\ref{SCIRS-model}) and accounting for (\ref{SCIRS-model_integral}) takes us to the particular simple 
form without memory
\beq
 \label{SCIRS-model_exponential_case}
 \begin{array}{clr}
\displaystyle  \frac{d}{dt}s(t) & = \displaystyle -{\cal A}(t) + \xi_R \, r(t)  & \\  \\
\displaystyle \frac{d}{dt}c(t) & = \displaystyle {\cal A}(t) -\xi_C \, c(t) \\ \\
\displaystyle \frac{d}{dt}j(t) & = \displaystyle \xi_C c(t) -\xi_I\, [j(t)-j_0]   \\ \\
 \displaystyle  \frac{d}{dt}r(t) & = \displaystyle \xi_I\, [j(t)-j_0] - \xi_R\, r(t) 
 \end{array} \hspace{1cm} (t\geq 0).
\eeq
Indeed exponential densities stand out by the memoryless feature and the Markov property (see textbooks, e.g. \cite{Tijms2003} and many others for details), thus
the transition rates in (\ref{SCIRS-model_exponential_case}) depend only on the instantaneous state $s(t),c(t),j(t),r(t)$ 
but not on the history of the evolution. 
Setting the transition rates on the left hand side to zero yields straight-forwardly the endemic state which we derive subsequently (Eqs. (\ref{general-endemic}), (\ref{gen_endemic_equilib})) for arbitrary waiting time kernels with existing means.
\subsection{Arbitrary waiting time distributions with memory}
\label{gen_memory}
To explore the general cases with memory consider now  the equivalent representations (\ref{SCIRS-model_integral}) and (\ref{SCIRS-model}) in the Laplace space
\begin{equation}
 \label{SCIRS-model-laplace}
 \begin{array}{clr}
\displaystyle  {\hat s}(\lambda) & = \displaystyle \frac{1-j_0}{\lambda} 
- {\hat {\cal A}}(\lambda)\frac{[1- {\hat K}_C(\lambda) {\hat K}_I(\lambda){\hat K}_R(\lambda)]}{\lambda} & \\  \\
\displaystyle {\hat c}(\lambda) & = \displaystyle {\hat {\cal A}}(\lambda)\frac{(1 - {\hat K}_C(\lambda)}{\lambda} \\ \\
\displaystyle {\hat j}(\lambda)  & = \displaystyle \frac{j_0}{\lambda} +   {\hat {\cal A}}(\lambda) {\hat K}_C(\lambda)\frac{(1 - {\hat K}_I(\lambda)}{\lambda}\\ \\
 \displaystyle {\hat r}(\lambda) & = \displaystyle {\hat {\cal A}}(\lambda) {\hat K}_C(\lambda){\hat K}_I(\lambda)\frac{(1 - {\hat K}_R(\lambda)}{\lambda}
 \end{array}
\end{equation}
where all Laplace transforms (LTs) depend on ${\hat {\cal A}}(\lambda)$ of the (unknown) infection rate and 
${\hat \Phi}_{C,I,R}(\lambda) = \frac{(1 - {\hat K}_{C,I,R}(\lambda)}{\lambda}$ are the LTs of the survival probabilities.
We further have that $${\hat s}(\lambda)+{\hat c}(\lambda)
+{\hat j}(\lambda)+{\hat r}(\lambda)=\frac{1}{\lambda} = \int_0^{\infty}e^{-\lambda t}{\rm d}t$$ as we deal with a constant population. 
We observe that in the first Eq. of (\ref{SCIRS-model-laplace})
the term
$\frac{[1- {\hat K}_C(\lambda) {\hat K}_I(\lambda){\hat K}_R(\lambda)]}{\lambda}$ is the LT of the survival probability
$\mathbb{P}[t_C+t_R+t_I > t]$ (due to the PDF $(K_c \star K_I \star K_R)(t)$ of the random variable $t_C+t_I+t_R$), corresponding to individuals `surviving' in one of the compartments C, I, R at time $t$.
To shed more light on how the memory comes into play we introduce a `memory operator'  (For our convenience we employ here a slightly modified definition as in  \cite{MainardiGorenfloScalas2004,MiPoRi2021}.)\footnote{We suppress here subscripts C, I, R.}
\beq
\label{memory}
{\hat {\cal M}}(\lambda) = \frac{{\hat K}(\lambda)}{{\hat \Phi}(\lambda)} = 
\lambda \frac{{\hat K}(\lambda)}{1-{\hat K}(\lambda)} = \lambda 
{\hat {\cal L}}(\lambda).
\eeq
Then reading 
$ {\hat K}(\lambda) = {\cal M}(\lambda) {\hat \Phi}(\lambda)$ as convolutions in the time domain we
can rewrite our SCIRS Eqs. (\ref{SCIRS-model}) (in the Caputo sense) of general (fractional) derivatives (see \cite{Kochubei2011,fractalfract2020,MiRia-FCAA2020} and Appendix) as
\begin{equation}
 \label{SCIRS-model_memory}
 \begin{array}{clr}
\displaystyle  \frac{d}{dt}s(t) & = \ds  - {\cal A}(t)  
+   \int_0^t{\cal L}_R(t-\tau) \frac{d}{d\tau} r(\tau){\rm d}\tau & \\  \\
\displaystyle \frac{d}{dt}c(t) & = \ds  {\cal A}(t) - \int_0^t{\cal L}_C(t-\tau) \frac{d}{d\tau} c(\tau){\rm d}\tau \\ \\
\displaystyle  \frac{d}{dt}j(t) & =\ds \int_0^t{\cal L}_C(t-\tau)\frac{d}{d\tau} c(\tau){\rm d}\tau  - \int_0^t{\cal L}_I(t-\tau)\frac{d}{d\tau}j(\tau){\rm d}\tau \\ \\
 \displaystyle  \frac{d}{dt}r(t) & =\ds  \int_0^t{\cal L}_I(t-\tau)
 \frac{d}{d\tau}j(\tau){\rm d}\tau - \int_0^t{\cal L}_R(t-\tau)\frac{d}{d\tau} r(\tau){\rm d}\tau.
 \end{array}
\end{equation}
These equations show that the transition rates at time $t$ have a complete memory of their previous values. Alternatively we can write (\ref{SCIRS-model_memory}) in the Riemann-Liouville manner of general fractional derivatives (see Appendix).
For exponential waiting time PDFs $K_{exp}(t)= \xi e^{-\xi t}$ we have
${\hat {\cal M}}_{exp}(\lambda) =  \xi$ thus ${\cal M}_{exp}(t) = \xi \delta(t)$
(i.e. is null for $t>0$ indicating lack of memory) and we easily recover above memoryless SCIRS Eqs. 
(\ref{SCIRS-model_exponential_case}).
\subsection{Long waiting times -- the time fractional case}
\label{time_fractional}
Representation (\ref{SCIRS-model_memory}) is especially useful when we deal with fat-tailed waiting time PDFs without existing means such as the Mittag-Leffler waiting-time PDF (ML-PDF) (see \cite{SamkoKilbas1993} and references therein for explicit formulas). The ML-PDF is a fractional generalization of the exponential PDF and has the LT \cite{MainardiGorenfloScalas2004,MiPoRi2021} (and many others)
\beq
\label{ML_density}
{\tilde K}_{\beta}(\lambda) = \frac{\xi}{\xi+\lambda^{\beta}} , \hspace{1cm} \beta \in (0,1), \hspace{0.5cm} \xi > 0
\eeq
where due to the fat tail the mean waiting time $-\frac{d}{d \lambda}{\tilde K}_{\beta}(\lambda)\big|_{\lambda=0} \to \infty$ does not exist as $\beta \in (0,1)$ (occurrence of extremely long waiting times). For $\beta=1$ (\ref{ML_density}) retrieves the LT of the exponential density.
Expanding the LT for $\lambda$ small ${\tilde K}_{\beta}(\lambda) \sim
1-\frac{\lambda^{\beta}}{\xi}$, Laplace inversion shows that ML-PDF has a fat power law tail $K_{\beta}(t) \sim \frac{\beta}{\Gamma(1-\beta)}\frac{t^{-\beta-1}}{\xi}$ ($t\to \infty$) corresponding to a long memory.
For the ML-case we have for (\ref{memory})
\beq
\label{fractional_ker}
{\hat {\cal L}}_{\beta}(\lambda) = \xi \lambda^{-\beta}
\eeq
where $\lambda^{-\beta}$ corresponds to the Riemann-Liouville fractional integral of degree $\beta$ and ${\hat {\cal M}}_{\beta}(\lambda) = 
\xi \lambda^{1-\beta}$ the Riemann -Liouville (R-L) fractional derivative of order $1-\beta$. Denoting $D_t^{\nu}$ the R-L fractional derivative where it is sufficient here to consider the range $\mu \in (0,1)$ defined by \cite{SamkoKilbas1993}
\beq
\label{RL_derivative}
D_t^{\mu} \cdot f(t) = \frac{d}{dt} \int_0^t \frac{(t-\tau)^{-\mu}}{\Gamma(1-\mu)}f(\tau){\rm d}\tau ,\hspace{2cm} \mu \in (0,1)
\eeq
having LT $\lambda^{\mu}{\hat f}(\lambda)$ 
and with the definition of the Caputo fractional derivative 
\beq
\label{Caputo_derivative}
\frac{d^{\mu}}{dt^{\mu}} f(t) =  \int_0^t \frac{(t-\tau)^{-\mu}}{\Gamma(1-\mu)}\frac{d}{d\tau} f(\tau){\rm d}\tau  = 
D_t^{\mu} \cdot f(t) - f(0)\frac{t^{-\mu}}{\Gamma(1-\mu)} ,\hspace{2cm} \mu \in (0,1)
\eeq
with LT $\lambda^{\mu}{\hat f}(\lambda)-f(0)\lambda^{\mu-1}$. Consult the Appendix
and \cite{Kochubei2011} for the connections with general (fractional) derivatives.
In the limit $\mu \to 1-$ both Caputo and RL-fractional derivatives converge to the first order standard derivative where $\frac{t^{-\mu}}{\Gamma(1-\mu)} \to \delta(t)$.
Using the feature
\beq
\label{Caputo_fearure}
\frac{d^{\mu}}{dt^{\mu}}[f(t)-f(0)] =\frac{d^{\mu}}{dt^{\mu}}f(t) = D_t^{\mu} \cdot [f(t)-f(0)]
\eeq
we can then write (\ref{SCIRS-model_memory}) when all waiting times $t_{C,I,R}$ are drawn from ML PDFs
in terms of Caputo fractional derivatives as
\begin{equation}
 \label{SCIRS-model_memory_time_fractional}
 \begin{array}{clr}
\displaystyle  \frac{d}{dt}s(t) & = \ds  - {\cal A}(t)  
+   \xi_R\frac{d^{1-\beta}}{dt^{1-\beta}}r(t) & \\  \\
\displaystyle \frac{d}{dt}c(t) & = \ds  {\cal A}(t) - 
\xi_C\frac{d^{1-\beta}}{dt^{1-\beta}}c(t)  \\ \\
\displaystyle  \frac{d}{dt}j(t) & =
\ds  \xi_C\frac{d^{1-\beta}}{dt^{1-\beta}}c(t)  - \xi_I\frac{d^{1-\beta}}{dt^{1-\beta}}j(t) \\ \\
 \displaystyle  \frac{d}{dt}r(t) & =\ds \xi_I\frac{d^{1-\beta}}{dt^{1-\beta}}j(t) - \xi_R\frac{d^{1-\beta}}{dt^{1-\beta}}r(t)
 \end{array}
\end{equation}
retrieving for $\beta=1$ the Eqs. 
(\ref{SCIRS-model_exponential_case}) of exponential waiting time PDFs (see (\ref{Caputo_derivative}), (\ref{Caputo_fearure})).
Indeed the time-fractional case with fat-tailed waiting time distributions is of utmost importance deserving further thorough investigation, see Appendix \ref{R0_finite_means} for a brief account.

\subsection{Endemic equilibrium for waiting time PDFs with existing mean}
\label{endemic_gen}
Here we confine ourselves to waiting time PDFs with existing means.
The endemic state is defined as the long-time limit of the evolution. That is we seek a stationary (constant solution)
$[s(t),c(t),j(t),r(t)] \to [S_e,C_e,J_e,R_e]$ and ${\cal A}(t) \to A_e=\beta J_eS_e$ which also is the asymptotic solution for $t \to \infty $ if the endemic state is stable. This solution can be obtained from Eqs.
(\ref{SCIRS-model-laplace}) in the limit of small $\lambda$, where these equations then take 
(with ${\hat s}(\lambda) \to S_e/\lambda$,..) the form 
\begin{equation}
 \label{SCIRS-model-laplace-lambda-to-zero}
 \begin{array}{clr}
\displaystyle  S_e & = \displaystyle   1-j_0 
-  \lim_{\lambda \to 0} {\hat {\cal A}}(\lambda) [1- K_C(\lambda) K_I(\lambda){\hat K}_R(\lambda)]  & \\  \\
\displaystyle  C_e & = \displaystyle \lim_{\lambda \to 0}  {\hat {\cal A}}(\lambda) [1 - {\hat K}_C(\lambda)] \\ \\
\displaystyle  J_e  & = \displaystyle   j_0 + \lim_{\lambda \to 0}  {\hat {\cal A}}(\lambda) {\hat K}_C(\lambda)[1 - {\hat K}_I(\lambda)] \\ \\
 \displaystyle  R_e & = \displaystyle  \lim_{\lambda \to 0}  {\hat {\cal A}}(\lambda) {\hat K}_C(\lambda){\hat K}_I(\lambda)[1 - {\hat K}_R(\lambda)]
 \end{array}
\end{equation}
with $${\hat {\cal A}}(\lambda) \approx \frac{A_e}{\lambda} + A_0+A_1\lambda + \ldots $$
which has to be considered for $\lambda \to 0$ where the lowest order in $\lambda$ determines the endemic equilibrium. 
Now with 
\beq
\label{Klamexp}
{\hat K}(\lambda) = \int_0^ {\infty}e^{-\lambda\tau} K(\tau){\rm d}\tau \approx 
\int_0^ {\infty}[1-\lambda\tau] K(\tau){\rm d}\tau +o(\lambda) 
= 1- \lambda \langle \tau \rangle +o(\lambda)
\eeq
where with $\langle ..\rangle$ we denote mean values and $o(\lambda)$ stands for the Landau symbol.
Thus we have
\beq
\label{Klamexpansions}
{\hat K}_{C,I,R}(\lambda) = 1 -\lambda \langle t_{C,I,R} \rangle +o(\lambda)
\eeq 
and ${\hat K}_{C}(\lambda){\hat K}_{I}(\lambda){\hat K}_{R}(\lambda) =1- \lambda 
\langle t_C+t_I+t_R \rangle +o(\lambda) = 1-\lambda \langle T \rangle +o(\lambda)$
and as said we assume that the means exist. Therefore,
$$
\lim_{\lambda \to 0}{\hat \Phi}_{C,I,R}(\lambda) =\lim_{\lambda \to 0} 
\frac{(1 - {\hat K}_{C,I,R}(\lambda)}{\lambda} = -\frac{d}{d\lambda}{\hat K}_{C,I,R}(\lambda)\big|_{\lambda=0} =
\langle t_{C,I,R} \rangle 
$$
Then we get straight-forwardly for the lowest orders in $\lambda$ the endemic equilibrium as follows
\beq
\label{general-endemic}
\begin{array}{clr}
\displaystyle  S_e & = \displaystyle 1-j_0
- A_e \, \langle T \rangle   & \\  \\
C_e & = \displaystyle A_e \, \langle t_C \rangle \\ \\
\displaystyle  J_e  & = \displaystyle j_0 +   A_e \langle t_I \rangle  \\ \\
 R_e & = \displaystyle A_e \,  \langle t_R \rangle
 \end{array}
\eeq
and with our assumption $A_e=\beta J_eS_e$ we arrive at
\beq
\label{gen_endemic_equilib}
\begin{array}{clr}
S_e(J_e) & = \displaystyle \frac{1-j_0}{1+\beta \langle\, T\, \rangle  J_e} & \\ \\
C_e(J_e) & = \displaystyle  \frac{(1-j_0)\beta \langle t_C \rangle J_e}{1+\beta \langle\, T\, \rangle J_e} \\ \\
J_e & = \displaystyle  j_0+ \beta \langle t_I \rangle J_e \frac{1-j_0}{1+\beta \langle T \rangle J_e} \\ \\
R_e(J_e) & = \displaystyle \frac{(1-j_0)\beta \langle t_R \rangle J_e}{1+\beta \langle\, T\, \rangle J_e}. & 
\\ & & 
\end{array}
\eeq
Consider now the third relation in (\ref{gen_endemic_equilib}) 
which is an implicit equation for
$J_e$ leading to
\beq
\label{Je_eq}
J_e^2-2aJ_e -b =0
\eeq
where we have introduced 
\beq
\label{a_b}
\begin{array}{clr}
\ds a &= \ds \frac{j_0}{2} + \frac{\beta \langle t_I \rangle (1-j_0)-1}{2\beta \langle T \rangle} = & \ds \frac{\langle t_R+t_C \rangle)j_0}{2 \langle T \rangle}+\frac{R_0-1}{2\beta \langle T \rangle} , \hspace{1cm} R_0=\beta \langle t_I \rangle \\ \\
\ds b &= \ds \frac{j_0}{\beta \langle T \rangle}. &
\end{array}
\eeq
The quantity $R_0=\beta \langle t_I \rangle $ can be interpreted as the `basic reproduction number' and is a crucial (control-) parameter where we come back to this interpretation subsequently. (\ref{Je_eq}) has the roots
\beq
\label{roots}
(J_e)_{1,2}= a \pm \sqrt{a^2+b}.
\eeq
Clearly the endemic equilibrium exists only for those roots with $J_e \in [0,1]$ where the full endemic state then is determined by (\ref{gen_endemic_equilib}).
For $b=0 $ ($j_0=0$) one root is zero corresponding to the healthy state $S_e=1$.
The second root is $J_e(j_0=0)= 2a= \frac{R_0-1}{\beta \langle T \rangle} \in [0, 1)$ for $R_0=\beta \langle t_I \rangle >1$ which therefore is the condition of the existence of an endemic state for $j_0=0$. In Appendix \ref{R0_finite_means} we show for waiting time PDFs with existing means that
$R_0>1$ also is the condition that the healthy state
is unstable thus an epidemic can start to spread. We also show there that the healthy state is always unstable if $K_I(t)$ has a fat tail (i.e. infinite mean).
Indeed for $\langle t_I \rangle \to \infty$ ($\beta, \langle t_C \rangle, \langle t_R \rangle$ kept finite) we have $R_0 \to \infty$ thus $J_e(j_0=0) \sim \langle t_I \rangle /\langle T \rangle \to 1$ and for $\beta \to \infty $ ($\langle t_I \rangle,\langle t_C \rangle, \langle t_R \rangle$ finite) we get
$J_e(j_0=0) \to \langle t_I \rangle/\langle T \rangle$.

\begin{figure}[H]
\centerline{\includegraphics[width=0.75\textwidth]{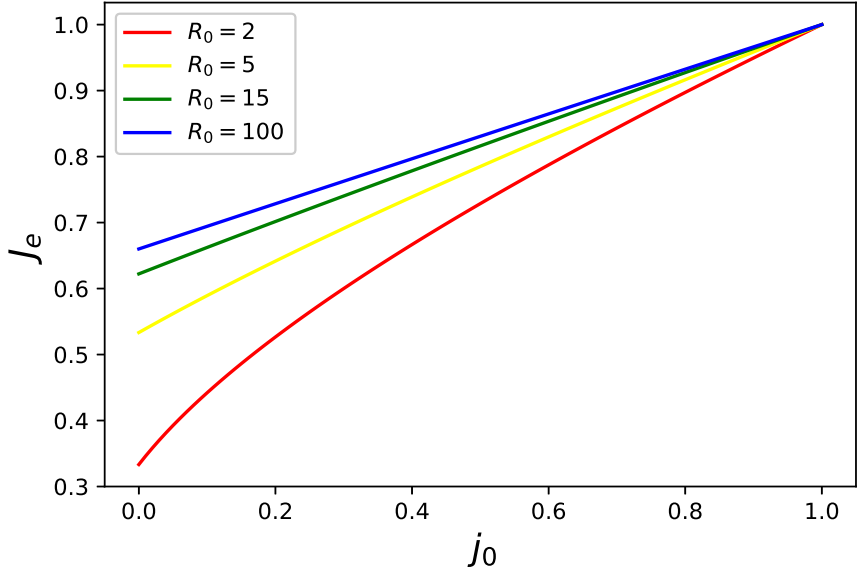}}
\caption{\label{Fig1} Endemic value 
$J_e(j_0,R_0)$ vs $j_0$ from Eq. (\ref{positiveroot}) with (\ref{a_b}) for different values of $R_0>1$ where in all curves we have put $\langle t_C \rangle =\langle t_R\rangle =5$, $\langle t_I\rangle =20$.}
\end{figure}
Consider now $j_0>0$ and $R_0>1$. Then we have $a <1$ as $\frac{\langle t_R+t_C\rangle j_0}{2T}< \frac{1}{2}$ and $\frac{R_0-1}{2\beta \langle T \rangle } <\frac{1}{2}$ and $b <1$
thus only the positive root is 
\beq
\label{positiveroot}
(J_e)_{1} =  
a + \sqrt{a^2+b} 
\eeq
whereas the other $(J_e)_{2}= a - \sqrt{a^2+b} < 0$.
The endemic state exists if $(J_e)_{1} \leq 1$. As $a,b$ are monotonously increasing functions of $j_0$ let us check the root for the maximum value
$j_0=1$ where
$a(j_0=1)=\frac{1}{2}-\frac{1}{2\beta T}=\frac{1}{2}(1-\epsilon)$ and $b(j_0=1) = \frac{1}{\beta \langle T \rangle }=\epsilon$.
Then we get
\beq
\label{Je1}
   J_{e}(j_0)\big|_{j_0=1} = \frac{1}{2}(1-\epsilon +\sqrt{(1-\epsilon)^2+4\epsilon} = \frac{1-\epsilon+\sqrt{(1+\epsilon)^2}}{2} = 1
\eeq
i.e. the initial and endemic state coincide $j_0=J_e(1)=1$ where we infer that
$j=1$ (for $R_0>1$) should be a stable endemic state.
From the monotony
of $J_e(j_0)$ (see (\ref{a_b})) we can see that 
$$ \frac{R_0-1}{\beta \langle T \rangle} \leq J_e(j_0) \leq J_{e}(j_0)\big|_{j_0=1} =1 $$
i.e. in the entire range $j_0 \in [0,1]$ the endemic value is given by the positive root (\ref{positiveroot}) together with (\ref{a_b}). This behavior is shown in the plot of Figure \ref{Fig1} where we draw $J_e$ versus $j_0$ for different values of $R_0>1$. One can see that $J_e$ is monotonously increasing with $R_0$ when $\langle t_R\rangle, \langle t_C \rangle, j_0$ are kept constant.

We focus now on initial conditions $s_0=1-$ and $j_0=0+$ (healthy state). 
The endemic equilibrium (\ref{gen_endemic_equilib}) then writes 
\beq
\label{endem}
\begin{array}{clr} 
\ds S_e &= \ds \frac{1}{R_0} & \\ \\
\ds C_e & = \ds \frac{R_0-1}{R_0} \frac{\langle t_C \rangle }{\langle T \rangle}  & \\ \\
\ds J_e & = \ds \frac{R_0-1}{R_0} \frac{\langle t_I \rangle}{\langle T \rangle}  &\\ \\
\ds R_e & = \ds \frac{R_0-1}{R_0} \frac{\langle t_R \rangle}{\langle T \rangle} & 
\end{array}  (\langle T \rangle = \langle t_C+t_I+t_R \rangle , \hspace{1cm} R_0=\beta \langle t_I \rangle)
\eeq
and exists solely for $R_0>1$ depending only from $R_0$ and the mean waiting times
$\langle t_{C,I,R}\rangle $.
We may consider the following limiting cases. 
\\ 
\noindent 1) $\langle t_R \rangle \to \infty $ (infinitely long immunity), $\langle t_C\rangle,\langle t_I\rangle ,\beta$ are kept constant: \\ 
Then we have $\frac{\langle t_R \rangle }{\langle T\rangle } \to 1$
 and 
$\frac{\langle t_C \rangle }{\langle T \rangle} , \frac{\langle t_I \rangle }{\langle T\rangle } \to 0$ and hence $S_e=\frac{1}{R_0}$ remains unchanged thus $C_e,J_e \to 0$ and
$R_e \to \frac{R_0-1}{R_0}$ with $S_e+R_e=1$ which corresponds to a fully healthy population (susceptible or immune) with $C_e+J_e \to 0$, i.e. the disease in the long time limit dies out.
\\ 
2) In the same way the limits
$ \langle t_I \rangle \to \infty$ (long illness time) and and long incubation time $\langle t_C \rangle \to \infty $, respectively, are straightforward, where the endemic values of respective compartments  with infinite waiting times
tend to $\frac{R_0-1}{R_0}$ with unchanged $S_e=\frac{1}{R_0}$. This corresponds to the fact that the individuals remain eventually trapped in the 
respective compartments (C or I) with infinite waiting times thus the cyclic transition pathway $S \to \ldots \to S $ is suppressed. 
We infer that in the limiting cases 1), 2) the endemic states are stable similar to the class of classical SIR models and do not exhibit oscillatory instabilities.
\section{$\delta$-distributed waiting times}
\label{delta_kernels}
An instructive case consists in the deterministic limit 
when the sojourn times in the compartments for all walkers 
are constant. Then the waiting time PDFs are Dirac $\delta$-functions
\beq
\label{sharp}
K_{C,I,R}(t)= \delta(t-t_{C,I,R})
\eeq
with LTs ${\hat K}_{C,I,R}(\lambda) = 
e^{-\lambda t_{C,I,R}}$.
Recall now the convolution of two $\delta$-kernels
$$\delta(t-t_1) \star \delta(t-t_2) = \delta(t-t_1-t_2)$$
which yields a new $\delta$-kernel
with a shifted peak at $t_1+t_2$. With this observation we have 
$(K_C\star K_I)(t) = \delta(t-t_C-t_I)$ and $(K_C\star K_I \star K_R)(t) = \delta(t-t_C-t_I-t_R)$. Then the SCIRS equations (\ref{SCIRS-model}) read
\beq
\label{SRIRS-delta}
\begin{array}{clr}
 \frac{d}{dt}s(t) & = -{\cal A}(t) +  {\cal A}(t-t_C-t_I-t_R) & \\ \\
\frac{d}{dt}c(t) & =  {\cal A}(t) -  {\cal A}(t-t_C) & \\ \\
\frac{d}{dt}j(t) & = {\cal A}(t-t_C) - {\cal A}(t-t_C-t_I) & \\ \\
\frac{d}{dt}r(t) & =  {\cal A}(t-t_C-t_I) -  {\cal A}(t-t_C-t_I-t_R). &
\end{array}
\eeq
Note that ${\cal A}(t)={\cal A}[j(t),s(t)]$ is causal, i.e. all functions with negative time arguments are vanishing. For $t<t_C$ the infected individuals accumulate in the compartment C and for  $t \geq t_C$ some start to leave compartment C to enter I with the delayed rate $A(t-t_C)$. Then for $t\geq t_C+t_I$
the first individuals are healed starting transitions I $\to$ R with
rate $A(t-t_C-t_I)$. Finally for $t\geq t_C+t_I+t_R$ transitions R $\to$ S occur due to individuals losing their immunity. 
From these observations
we can infer that 
\beq
\label{start_j}
\begin{array}{clr}
c(t)= & \displaystyle  \int_0^t{\cal A}(\tau)\,{\rm d}\tau - \, \Theta(t-t_C)\int_0^{t-t_C} {\cal A}(\tau)\,{\rm d}\tau &  \\ \\
j(t)   =  &\displaystyle j_0 + \,\Theta(t-t_C)\int_0^{t-t_C} {\cal A}(\tau)\,{\rm d}\tau \, -  \Theta(t-t_C-t_I)\int_0^{t-t_C-t_I} {\cal A}(\tau)\,{\rm d}\tau& \\ \\
r(t) = & \displaystyle \Theta(t-t_C-t_I)\int_0^{t-t_C-t_I} {\cal A}(\tau)\,{\rm d}\tau - \,
 \Theta(t-t_C-t_I-t_R)\int_0^{t-t_C-t_I-t_R} {\cal A}(\tau)\,{\rm d}\tau
\end{array}
\eeq
where $\Theta(\tau)$ denotes the Heaviside unit step function defined as $\Theta(\tau)=1$ for $\tau\geq 0$ and $\Theta(\tau)=0$ for $\tau<0$. We can see for sharp waiting times the Heaviside functions `switch on and off' the respective transitions between compartments. It is worthy of mention that
(\ref{start_j}) is consistent with (\ref{SCIRS-model_integral})
when we 
take into account the survival probabilities 
\beq
\label{survival_comp}
\Phi_{C,I,R}(t) = \int_t^{\infty} \delta(\tau-t_{C,I,R}){\rm d}\tau =
\Theta(t_{C,I,R}-t)= 1-\Theta(t-t_{C,I,R})
\eeq
i.e. $\Phi_{C,I,R}(t) = 1$ for $t<t_{C,I,R}$ (`survival' in C, I, R, respectively) and 
$\Phi_{C,I,R}(t) = 0$ for $t>t_{C,I,R}$ (`death', having left C, I, R, respectively).
From Eqs. (\ref{start_j}) we find the endemic equilibrium $S_e,C_e,J_e,R_e$ representing a stationary solution of
(\ref{SRIRS-delta}). Plugging the stationary value ${\cal A}_e=\beta J_e S_e$
into (\ref{start_j}) we re-arrive at Eqs. (\ref{gen_endemic_equilib}) with
$t_{C,I,R} = \left\langle t_{C,I,R} \right\rangle$ for Dirac $\delta$-distributions. 

\subsection{Stability analysis of endemic and healthy states}
\label{R0}
Here we investigate the stability of the endemic equilibrium for healthy initial conditions $s_0=1$ ($j_0=0$) for $\delta$-distributed waiting times. 
To this end we set
\beq
\label{stability_analysis}
\begin{array}{clr}
s(t) = & S_e + u e^{\mu t} & \\ \\
c(t)=  & C_e + v e^{\mu t} & \\ \\ 
j(t) = & J_e + w e^{\mu t} &  \\ \\
r(t) = & R_e + x e^{\mu t}
\end{array}
\eeq
where $u,v,w,x$ are `small' time independent constants. Clearly at least one of these equations is redundant as $s+c+j+r=1$.
Then it follows that
${\cal A}(t) \approx A_e + A_0(u,w) e^{\mu t}$ where we take into account only the linear orders in
$u,v,w,x$. Therefore,
$A_e=\beta J_e S_e$ and $A_0(u,w)= \beta(u J_e+w S_e)$.
Plugging this into (\ref{SRIRS-delta}) leads to the following system of equations
\beq
\label{cond_eqs}
\begin{array}{clr}
\ds  A_0(u,w)(1-e^{-\mu T})+\mu u & =0 & \\ \\
\ds  A_0(u,w)(1-e^{-\mu t_C}) - \mu v & = 0 & \\ \\
\ds  A_0(u,w)e^{-\mu t_C}(1-e^{-\mu t_I}) - \mu w & =0 &  \\ \\
\ds   A_0(u,w)e^{-\mu (t_C+t_I)}(1-e^{-\mu t_R}) - x\mu  & = 0. &
\end{array}
\eeq
We can eliminate $v=A_0(u,w)\frac{1-e^{-\mu t_C}}{\mu}$ and $x= A_0(u,w)e^{-\mu (t_C+t_I)} \frac{1-e^{-\mu t_R}}{\mu}$ which are uniquely determined by $u$ and $w$. Therefore the solvability
condition
is determined uniquely by the first and third equation for $s$ and $j$ containing only 
the coefficients $u$ and $w$. The complete determinant of the system (\ref{cond_eqs}) leads to
\beq
\label{det_solv}
\left\| \begin{array}{cl}\ds  \beta J_e(1-e^{-\mu T})+\mu ;& \ds \beta S_e(1-e^{-\mu T}) \\ \\   \ds \beta J_e e^{-\mu t_C}(1-e^{-\mu t_I}) ; & \ds \beta 
 S_e 
e^{-\mu t_C}(1-e^{-\mu t_I}) -\mu \end{array} \right\| \mu^2 = 0.
\eeq
We have $\mu=0$ as a threefold eigenvalue
and with one non-zero eigenvalue $\mu_1$ determined by 
\beq
\label{eigenval}
\mu= \beta S_e e^{-\mu t_C}(1-e^{-\mu t_I})- \beta J_e(1-e^{-\mu T})
\eeq
and hence $\mu_1=\mu_1(\beta,j_0,t_C,t_I,t_R)$. However, we will see that we can reduce the set of pertinent parameters.
For the outbreak of an epidemic, it is necessary that the healthy (initial)
state $s_0=1$ becomes unstable. To explore this issue we
consider $s(t)=s_0+u e^{\mu t}$ and $j(t)=j_0+we^{\mu t}$ with $s_0=1$, $j_0=0$. 
Then we get the solvability condition by replacing in (\ref{det_solv}) $J_e \to j_0=0$ and $S_e \to s_0=1$ in the form
\beq
\label{stability_healthy}
 {\tilde \mu} = R_0e^{-{\tilde \mu}t_1}[1-e^{-{\tilde \mu}}) =  g({\tilde \mu},R_0)
\eeq
where we introduce
\beq
\label{h-relat}
\begin{array}{clr}
\ds {\tilde \mu}= t_I \mu , & \ds  t_1=\frac{t_C}{t_I} , & \ds t_2=\frac{T}{t_I}=1+t_1+\frac{t_R}{t_I}.
\end{array}
\eeq
The healthy state becomes unstable if there is a positive solution
${\tilde \mu}_0>0$ of this equation. Clearly there is a positive root only
if $\frac{d}{d{\tilde \mu}}g({\tilde \mu},R_0)|_{{\tilde \mu}=0}= R_0>1$, otherwise the healthy state is stable. This justifies our interpretation
of $R_0$ as the basic reproduction number 
where the healthy state is unstable for $R_0>1$ thus the epidemic starts to spread. 

\begin{figure}[H]
\centerline{\includegraphics[width=0.75\textwidth]{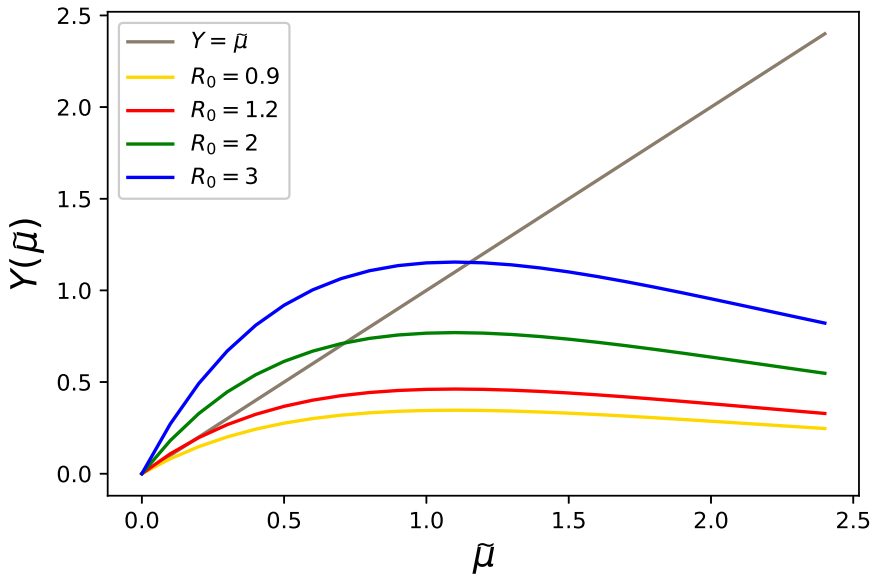}}
\caption{\label{Fig0} We depict $g({\tilde \mu},R_0)= R_0e^{-{\tilde \mu}t_1}(1-e^{-{\tilde \mu}})$ for different values of $R_0$. For $R_0=0.9$ (lower curve) the healthy state is stable. In the other curves $R_0>1$ the healthy state is unstable. In all plots we chose $t_1=t_C/t_I= 0.5$.}
\end{figure}

It is instructive to connect this interpretation with the common definition of $R_0$. The (dimensionless) basic reproduction number $R_0$ is defined as the expected number of infections caused by one infectious individual in a healthy (susceptible) population. 
We get this information when we multiply the second equation of (\ref{SCIRS-model}) with the total number of individuals $Z$ where $Z {\cal A}(t)$ then indicates the number of new infections per time unit. Considering this quantity at $t=0$ for the initial condition $j_0=\frac{1}{Z}$ and $s_0=1-j_0$
(initial condition of one infected individual $Z_I(0)=j_0 Z= 1$ in a healthy, i.e. susceptible population $Z_S(0)=s_0 Z=Z-1$), this takes us to
$$
\frac{dZ_c(t)}{dt}\big|_{t=0} = Z \beta s(t)j(t)\big|_{t=0} = \frac{\beta}{Z}Z_S(t)Z_I(t)\big|_{t=0} = \beta \frac{Z-1}{Z} \to \beta ,\hspace{1cm} Z \gg 1
$$
which is the number of new infections per time unit at $t=0$. Assuming this rate being constant during the average period of infection $\langle t_I\rangle$ (given this mean exists) yields the average number of new infections caused by the first infected walker during the period $t_I$ of his infection (in a susceptible population) as $R_0=\beta \langle t_I\rangle$. This is indeed the exact result of our model as we show in 
Appendix \ref{R0_finite_means}. We demonstrate there that the condition for an outbreak (condition of instability of the healthy initial state) is $R_0=\beta \langle t_I \rangle >1$.
This can also be seen more closely
in the right-hand side of (\ref{stability_healthy}) where $g({\tilde \mu},R_0)$ is a concave function of ${\tilde \mu}$ (see Figure \ref{Fig0}). 
We have for ${\tilde \mu}$ small the expansion $g({\tilde \mu},R_0) = R_0{\tilde \mu} + O({\tilde \mu}^2) > {\tilde \mu}$ only if $R_0>1$ and always $g({\tilde \mu},R_0) \to  0 < {\tilde \mu}$ as $\mu \to \infty$. Therefore exists a positive root $\mu_0(R_0)>0$ only for $R_0>1$. This behavior is sown in
Figure \ref{Fig0} where $\mu_0(R_0)$ increases monotonically with $R_0$ enhancing the (non-persistent) exponential growth of $j(t)$ at the outbreak of the epidemic.
\noindent Returning to the stability analysis of the endemic state we re-scale (\ref{eigenval}) with (\ref{h-relat}) as
\beq
\label{rescaled_egval}
{\tilde \mu} = R_0 S_e e^{-{\tilde \mu}t_1}(1-e^{-{\tilde \mu}}) 
- R_0 J_e(1-e^{-{\tilde \mu}t_2}).
\eeq
At threshold of an oscillatory instability, the eigenvalue is purely imaginary.
Plugging $\tilde\mu=i\omega$ into (\ref{rescaled_egval}) and separating real and
imaginary parts yields the two conditions
\beq
\label{Hopf_cndition}
\begin{array}{clr} 
f_1 (r_0,\omega) & = \cos\omega t_1- \cos\omega (t_1+1) -r_0(1-\cos\omega t_2) & = 0  \\ \\ 
f_2(r_0,\omega) & = \omega + \sin\omega t_1 - \sin\omega (t_1+1) +r_0\sin\omega t_2  & =  0
\end{array}
\eeq
which need to be both simultaneously fulfilled for an oscillatory (Hopf) instability where we introduced the reduced control parameter $r_0=J_e R_0 = (R_0-1)/t_2$.
\begin{figure}[H]
\centerline{\includegraphics[width=1.0\textwidth]{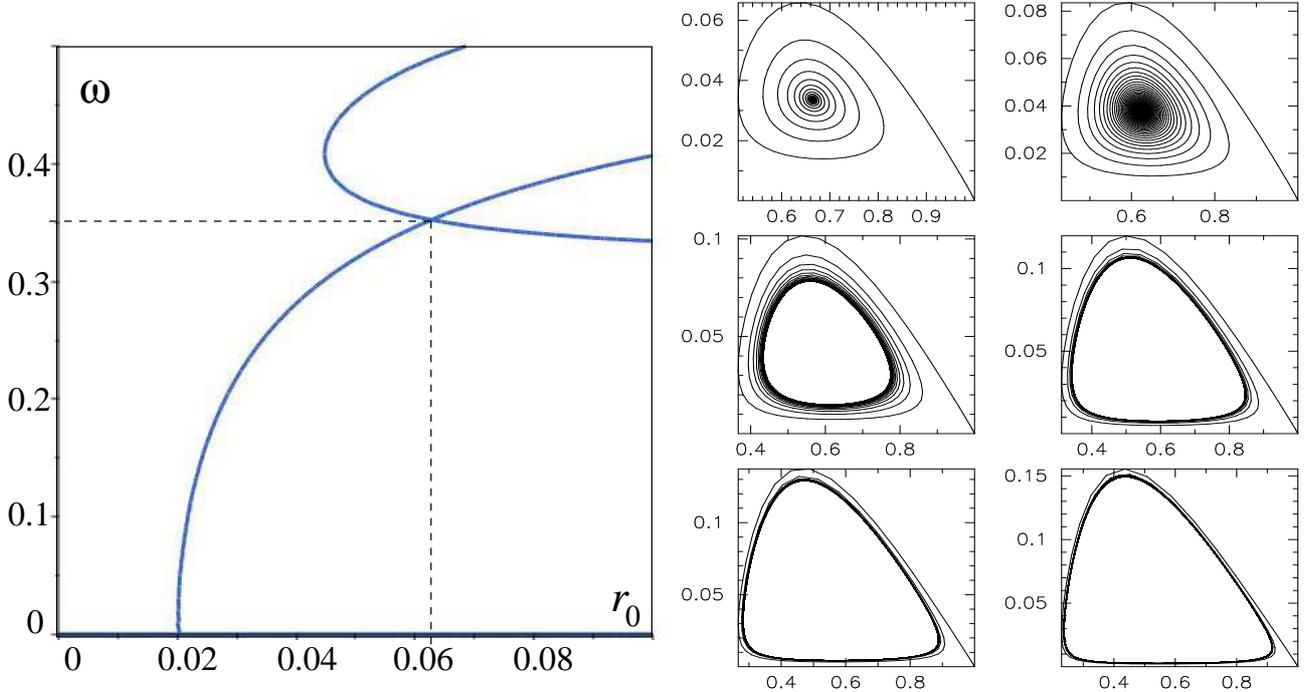}}
\caption{\label{linear} Left: the intersection of the contour lines of
$f_1,f_2=0$ of (\ref{Hopf_cndition}) yield $(\omega,r_0) \approx (0.35,0.063)$. Right:
  numerical solutions of the fully nonlinear delay system (\ref{SRIRS-delta}) for
  $R_0=1.5$ (top left) to $R_0=2.0$ in the $(j,s)$-plane. A limit cycle is born at
  $R_0\approx1.63$.}
\end{figure}
\noindent In Figure \ref{linear} (left frame) we plot the zero
lines of $f_i$ in the $r_0,\omega$ plane for $t_1=1/2,\ t_2=10$. They intersect at
$r_0\approx0.063,\ \omega=0.35$, corresponding to a critical $R_0\approx1.63$.
The right frame of Figure \ref{linear} shows numerical solutions of the full delay system
(\ref{SRIRS-delta}) with the same parameters $t_1,\ t_2$. The system is solved
using an Euler forward scheme with fixed time step $\delta t=10^{-4}$. The initial
conditions are $j_0=10^{-4},\ s_0=1-j_0,\ r_0=c_0=0$ close to the healthy state. We varied
$R_0$ from 1.5 (subcritical, upper left) to 2.0 (lower right frame). Clearly for $R_0=1.7$ the
stable focus turns into a limit cycle that becomes wider with increasing $R_0$.

\section{Microscopic model and computer simulations}
\label{C_python}
\subsection{Simple random walk model}
\label{simple_walk}
To explore the SCIRS phenomenology 
we combine this model with a multiple random walker's approach which we implemented into a
Python code \cite{supplementary-mat}.
In this random walk $Z$ walkers navigate independently on a periodic two dimensional lattice.
Each walker performs at integer times $t=0,1,2,\ldots$ instantaneous independent random steps to a closest neighbour lattice point (simple walk).
The position of walker $j$ ($j=1,\ldots Z$) can be described by the random variables
\beq
\label{random_walk}
\begin{array}{clr}
x_j(t) & = x_j(t-1) + \eta_x^{(j)}(t) & \\ \\
y_j(t) & = y_j(t-1) + \eta_y^{(j)}(t) &
\end{array} , \hspace{1cm} t=1,2,\ldots 
\eeq
with the random steps
$$
\left(\eta_x^{(j)}(t),\eta_y^{(j)}(t)\right) = (1,0); \, (-1,0); \, (0,1); \, (0,-1) 
 \hspace{0.25cm} {\rm occurring} \hspace{0.25cm} {\rm with} \hspace{0.25cm} {\rm probability} \hspace{0.25cm} \frac{1}{4}.
$$
This simple multi-walkers motion is a microscopic model with scaling limits to (standard) Brownian motion \cite{Feller1968}.
We assume a
$N_x=N_y=N$-periodic lattice with
$x^{(j)}(t) = x^{(j)}(t)\,\, {\rm mod}\,\, N$, $y^{(j)}(t)= y^{(j)}(t) \,\, {\rm mod} \,\, N$
for the position of each walker $j$.
In order to connect the random walk with the epidemic dynamics
we apply the following infection rule \cite{besmi}:
If a walker $j \in $ I meets a walker $k \in $ S on the same lattice point (`collision' of an infected--infectious and a susceptible walker), i.e.
\beq
\label{infection_rule}
\begin{array}{clr}
 \,\,
j \in I & \land \hspace{0.2cm} k \in S  &\\ \\
\land & x_j(t)  =  x_k(t) & \\ \\
\land & y_j(t)  =  y_k(t) & \\ \\
\end{array}
\eeq
then the susceptible 
walker $k$ gets infected with probability 
$P_{inf}$ and undergoes an
instantaneous transition S $\to$ C followed by the transition pathway as described previously. The infection probability $P_{inf}$ is constant for all walkers and time independent.
The random paths of the walkers are not affected by transitions between the compartments or 'collisions' of walkers. 
In the simulations at each integer time instant $t$ we count
the populations $Z_{S,C,I,R}(t)$ in the compartments where the total population $Z=Z_S(t)+Z_C(t)+Z_I(t)+Z_R(t)$ remains constant over time. As in the macroscopic model we focus on the fractions $s(t)=Z_S(t)/Z,\, c(t)=Z_C(t)/Z,\, j(t)=Z_I(t)/Z,\, r(t)=Z_R(t)/Z$. 

We implement the mutually independent waiting times $t_{C,I,R}$ as random numbers drawn from specific Gamma PDFs
$K_{C,I,R}(t)$. The Gamma-distribution
provides sufficient flexibility to generate a wide range of possible behaviors such as
sharp $\delta$-peaks, broadly scattered waiting times with a maximum or a monotonically decreasing PDF.
The Gamma (also called Erlang-) PDF writes 
\beq
\label{Gamma-dis}
K_{\xi,\alpha}(\tau) = \frac{\xi^{\alpha}\tau^{\alpha-1}}{\Gamma(\alpha)}e^{-\xi\tau} ,\hspace{1cm} \alpha, 
\xi   > 0
\eeq
where $\alpha$ is the so-called shape parameter and $1/\xi$ the time scale parameter. The Gamma PDF has a maximum for 
$\alpha > 1$ at $t_m=\frac{\alpha-1}{\xi}$ and is monotonically decreasing for $\alpha \leq 1$ and weakly singular at $t=0$ for $\alpha<1$. 
For $\alpha=1$ we get the exponential PDF $K_{\xi,1}(\tau)= \xi e^{-\xi\tau}$.
Some cases of Gamma PDFs are drawn in histograms of Figure \ref{Figure-B}.
Useful is its Laplace transform
\beq
\label{Laplace}
{\hat K}_{\xi,\alpha}(\lambda) = \int_0^{\infty} e^{-\lambda \tau} K_{\xi,\alpha}(\tau){\rm d}\tau = 
\frac{\xi^{\alpha}}{(\lambda+\xi)^{\alpha}}
\eeq
from which we can easily retrieve \\ ${\hat K}_{\xi,\alpha}(\lambda)\big|_{\lambda=0}=1$ (normalization), $ \left\langle t\right\rangle =
-\frac{d}{d \lambda}{\hat K}_{\xi,\alpha}(\lambda)\big|_{\lambda=0} = \frac{\alpha}{\xi}$ (mean waiting time),
and \\ ${\cal V} = \langle t^2 \rangle - \left\langle t\right\rangle^2 = \frac{\alpha}{\xi^2}$ (variance).
We use in the simulations to generate constant (sharp) waiting times the feature, e.g. \cite{besmi}
\beq
\label{Gamma-dis-deltalimit}
\lim_{\xi\to \infty} K_{\xi,\alpha=\xi \tau_0}(\tau) = \delta(t-\tau_0)
\eeq
where the mean $\tau_0 = \langle \tau \rangle $ is kept constant in this limit and the variance is vanishing. 
This limit is easily seen in the Laplace space
\beq
\label{Gamma-dis-deltalimit-laplace}
\lim_{\xi\to \infty} 
{\hat K}_{\xi,\xi \tau_0}(\tau) = \lim_{\xi \to \infty} \left(1+\frac{\lambda}{\xi}\right)^{-\xi \tau_0} = 
e^{-\tau_0 \lambda} = \int_0^{\infty} e^ {-\lambda t} \delta(t-\tau_0) {\rm d}t .
\eeq
\subsection{Validation of the macroscopic SCIRS model and case study}
\label{case_study}

In the simulations we remove unimportant fluctuations by recording the ensemble averaged\footnote{We denote here ensemble averages of random functions $B(t)$ with $\langle B(t) \rangle$.} compartment populations
$\langle s(t)\rangle ,\langle c(t)\rangle ,\langle r(t)\rangle ,
\langle j(t) \rangle$. We average numerically over a number of equivalent random walk realizations with identical parameters, waiting time distributions and observation times. Each realization employs different random numbers (`python seeds') for the waiting times and random walk.
We perform a case study to validate the macroscopic Eqs. (\ref{endem}) for the endemic equilibrium which exists for $R_0=\beta \langle t_I \rangle > 1$. We realize `natural' initial conditions close to the healthy state with one infected walker (in compartment I) and $Z-1$ susceptible walkers at $t=0$ (start of the experiment). 
In all computer experiments the walkers have random initial positions on the lattice. We determine the endemic equilibrium values numerically
by using the asymptotic relation
\beq
\label{S_e}
 [S_e,C_e,J_e,R_e]_{num}(t) \approx \frac{1}{t} \sum_{r=1}^t \langle [s(r),c(t),j(t),r(t)] \rangle
\eeq
converging for $t\to \infty$ to the endemic state $S_e,C_e,J_e,R_e$ if it exists. 
We measure the accordance of our macroscopic endemic equations (\ref{endem}) with the random walk approach by computing the ratios of the numerically determined endemic values and the values computed with (\ref{endem})
\beq
\label{model_ratios}
\begin{array}{clr}
r_{C} & = \frac{(C_e)_{num}}{[1-(S_e)_{num}]\langle t_C \rangle /\langle T\rangle} & \\ 
r_{I} & = \frac{(J_e)_{num}}{[1-(S_e)_{num}]\langle t_I \rangle /\langle T\rangle} &  \\ 
r_{R} & = \frac{(R_e)_{num}}{[1-(S_e)_{num}] \langle t_R \rangle /\langle T\rangle}
\end{array}
\eeq
by employing the numerically determined $S_e$.
In all simulations with existing endemic equilibria and sufficiently large observation time the ratios 
$r_{C,I,R} \approx 1+ O(10^{-2})$ are up to a few percent close to one, 
confirming impressively the prediction $C_e : J_e : R_e = \langle t_C \rangle : \langle t_I \rangle :
\langle t_R \rangle$ of the macroscopic model Eqs. (\ref{endem}). 
In the following discussion we give numerical evidence that our macroscopic SCIRS model well suits with the microscopic random walk approach. 
\begin{figure}[H]
\vspace{-0.5cm}
\includegraphics[width=0.52\textwidth]{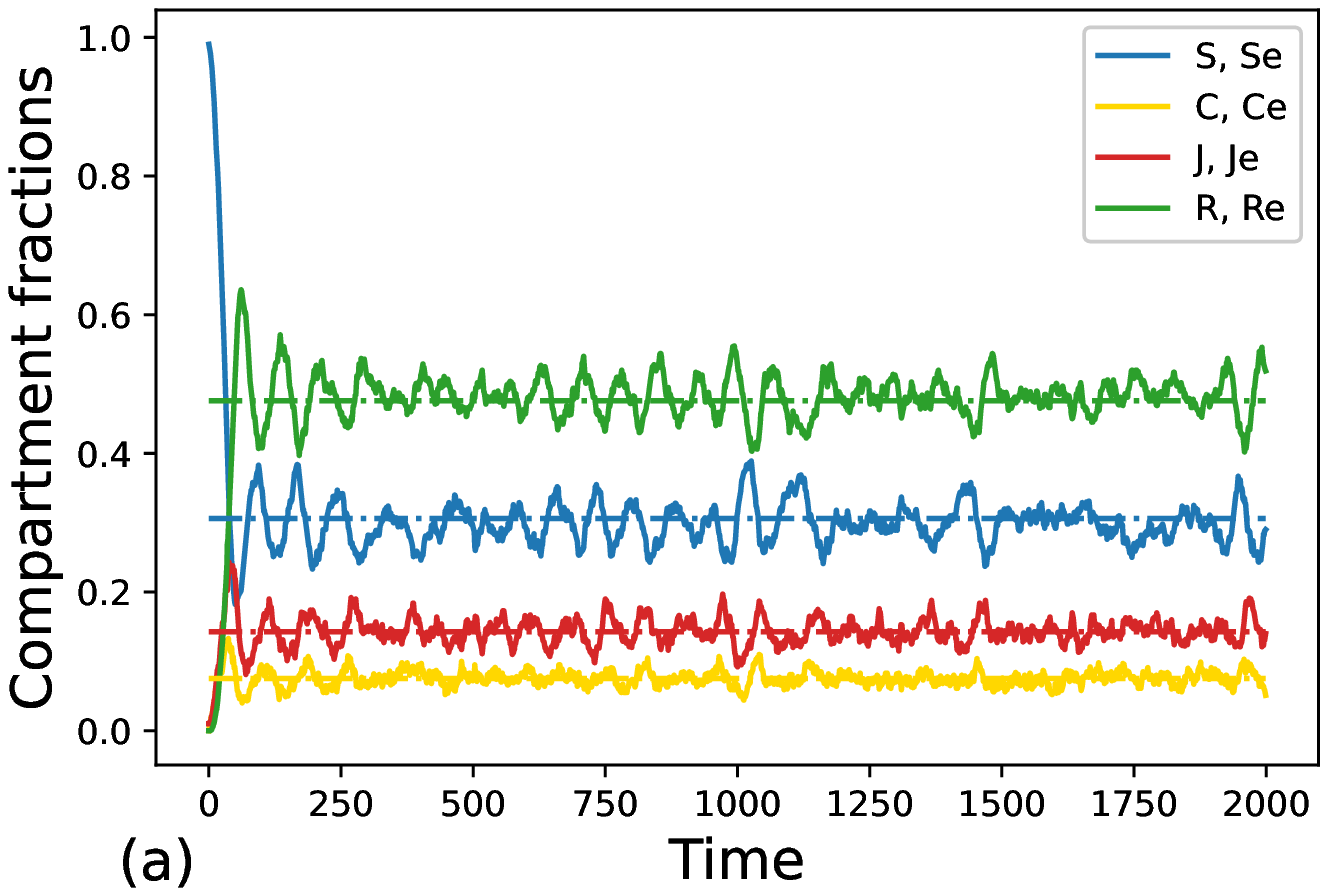}
\includegraphics[width=0.52\textwidth]{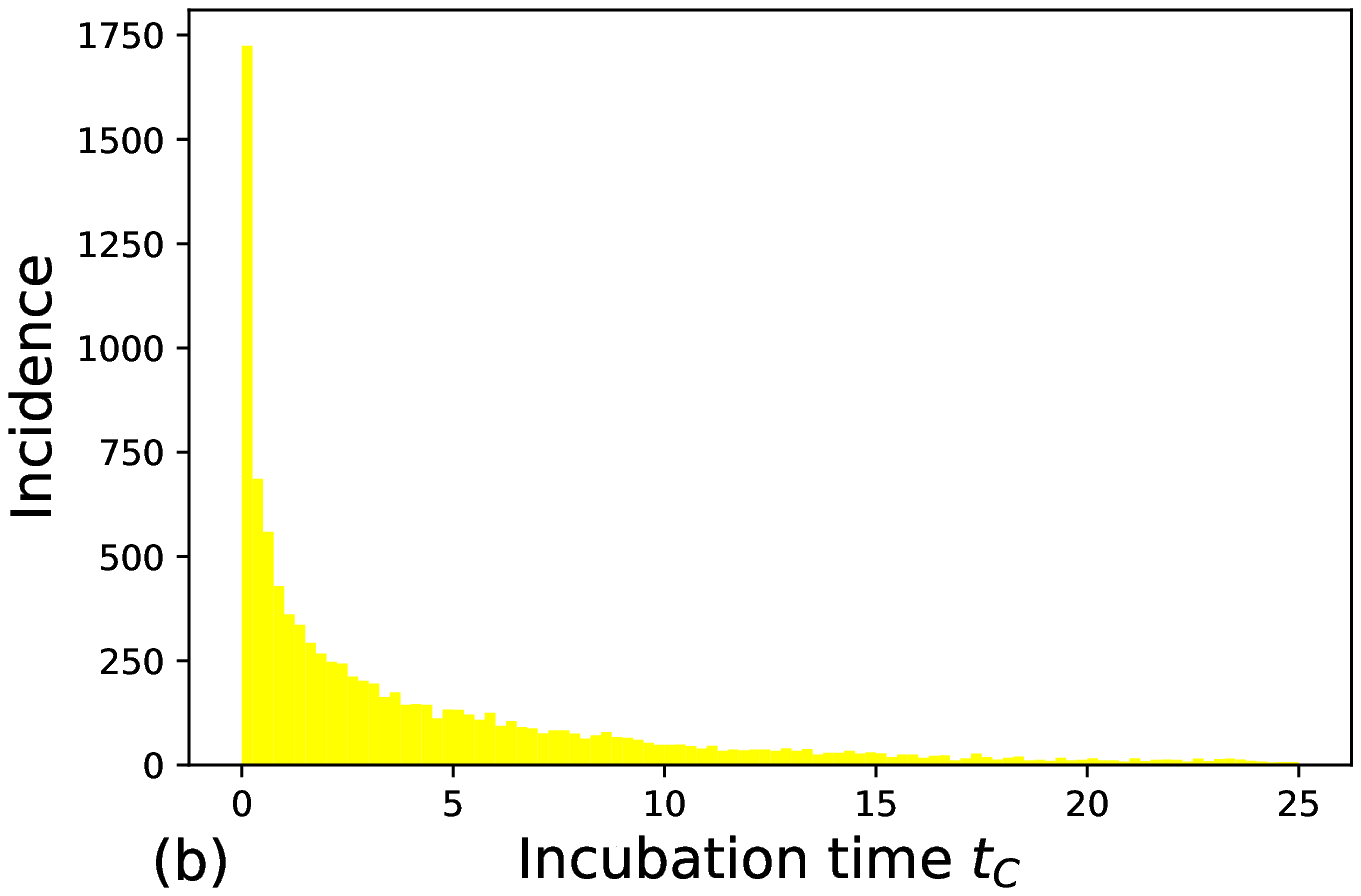}
\includegraphics[width=0.52\textwidth]{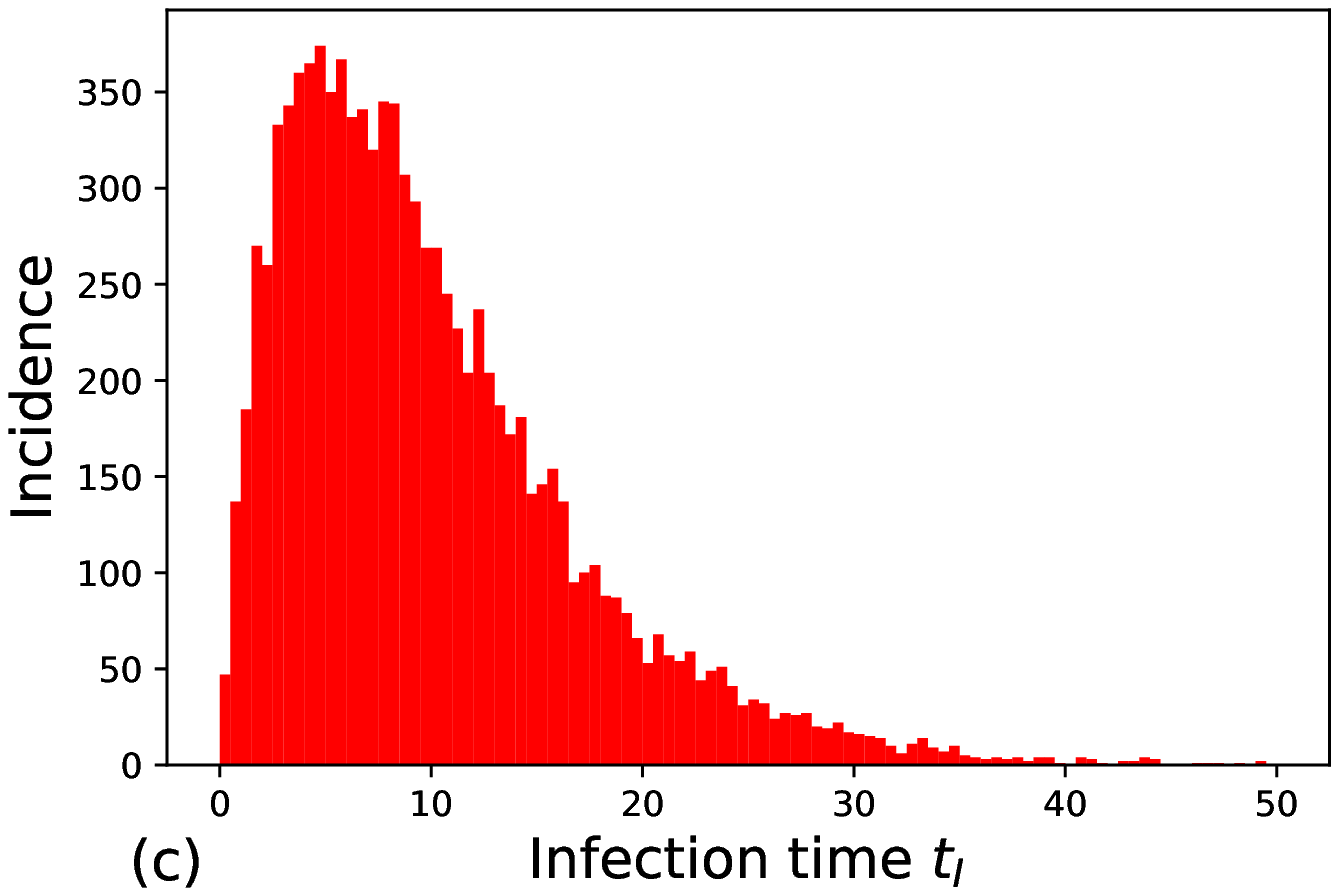}
\includegraphics[width=0.52\textwidth]{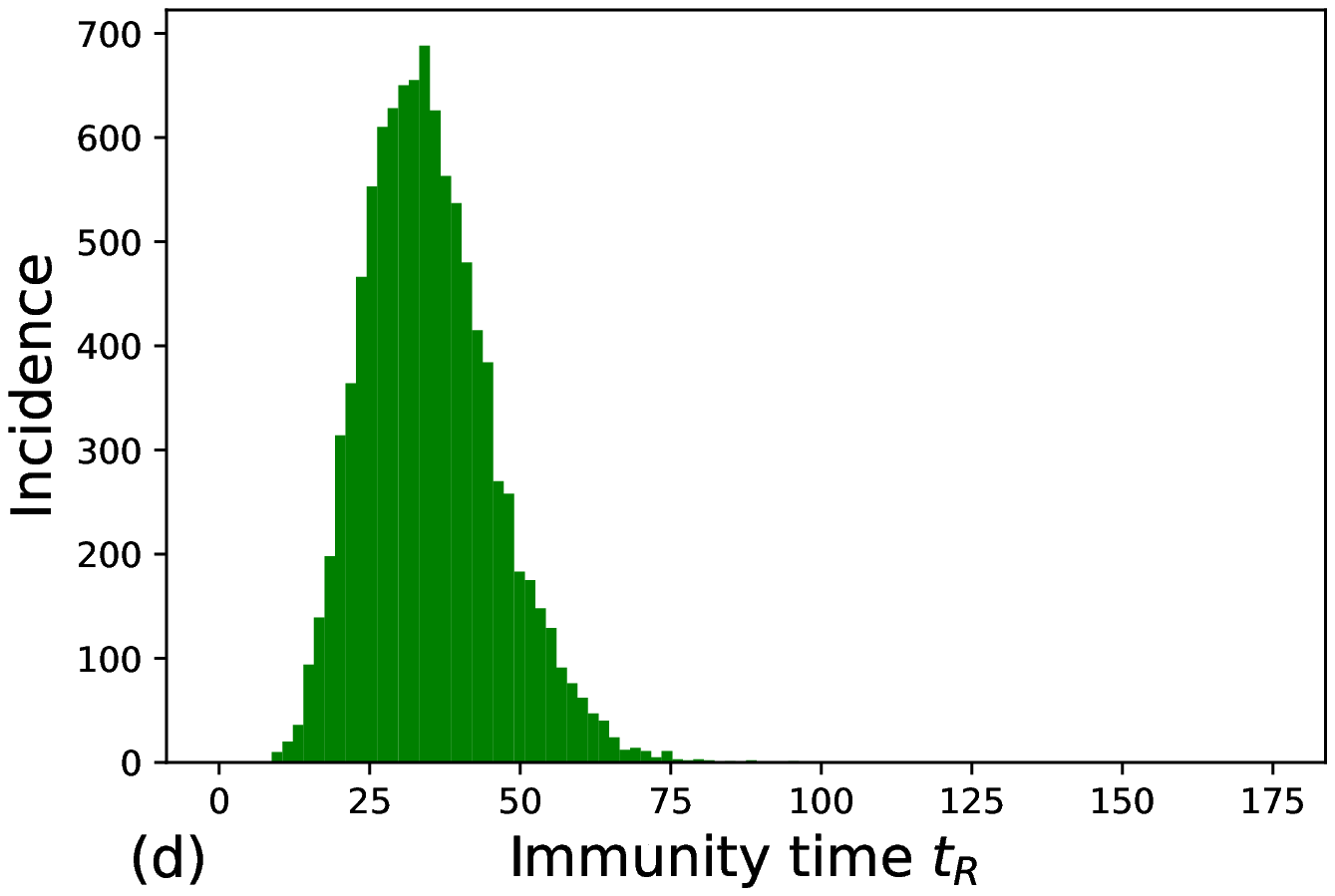}
\caption{\label{Figure-B} Compartment fractions averaged over $10$ random walk realizations
with $Z=100$, $N=11$ (density $Z/N^2 \approx 0.83$), $P_{inf} = 0.9$
and Gamma distributed waiting times having the means $\langle t_C \rangle = 5$, $\langle t_I \rangle = 10$, $\langle t_R \rangle = 35$, $\xi_C=0.1$, $\xi_I=0.2$, $\xi_R=0.3$.
By using Eq. (\ref{S_e}) for the numerical evaluation we get (dashed lines) $S_e \approx 0.31$,  $C_e \approx 0.075$, 
$J_e \approx 0.14$, $R_e \approx 0.48$ and with Eq. (\ref{model_ratios}) we have
$r_C \approx 1.08$, $r_I \approx 1.03$, $r_R\approx 0.98$. We depict the corresponding Gamma distributed waiting times $t_{C,I,R}$ in three histograms (b), (c), (d) with the parameters and color code used in the simulation of (a).
}
\end{figure}
\noindent In the experiment of Figure \ref{Figure-B} the waiting times are broadly scattered
and distributed by Gamma distributions of different parameters. For the chosen parameters the epidemic dynamics converges rapidly to the endemic states
where the ratio $C_e : J_e : R_e = \langle t_C \rangle : \langle t_I \rangle :
\langle t_R \rangle$ predicted by Eqs. (\ref{endem}). 
\begin{figure}[H]
\centerline{\includegraphics[width=0.75\textwidth]{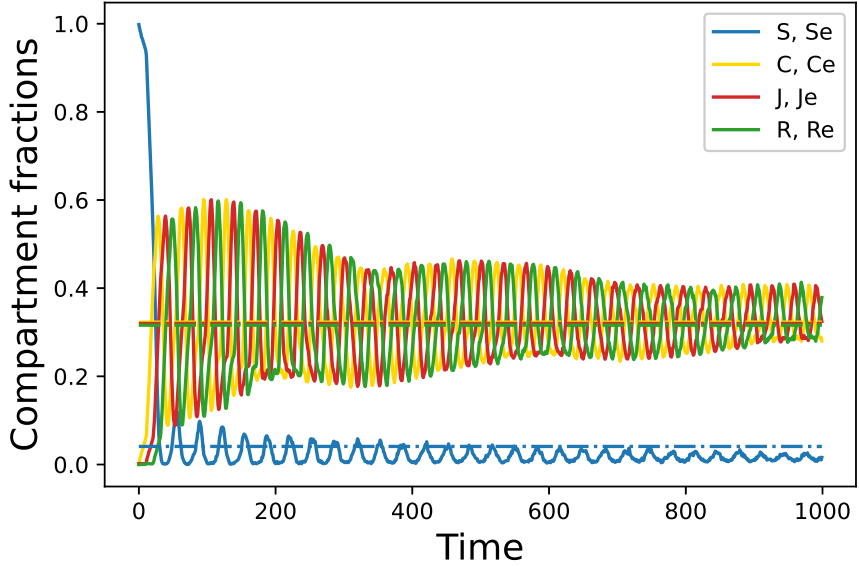}}
\caption{\label{coherent_moy} (a) Average over $5$ random walk realizations and ($\delta$-distributed) waiting times $t_C=t_I=t_R=10$ ($T=30$)
with $Z=500$, $N=11$ (density $Z/N^2 \approx 4.13$), $P_{inf} = 0.9$. 
We have small $S_e\approx 0.04$ ($R_0 \approx 24,39$) and $C_e=J_e=R_e \approx 0.32 \approx \frac{1}{3}$ (dashed lines).}
\end{figure}
\noindent In Figure \ref{coherent_moy} we have equal $\delta$-distributed waiting times and very large $R_0$. The measures (\ref{model_ratios}) are $r_{C,I,R} \approx (1.01, 1.00, 0.99)$ close to one and indicate excellent accordance 
with Eqs. (\ref{endem}) although the observation time is not very large. In this plot (blue dashed line) 
$S_e$ is slightly overestimated as (\ref{S_e}) is an asymptotic relation holding for large $t$.
\begin{figure}[H]
\includegraphics[width=0.52\textwidth]{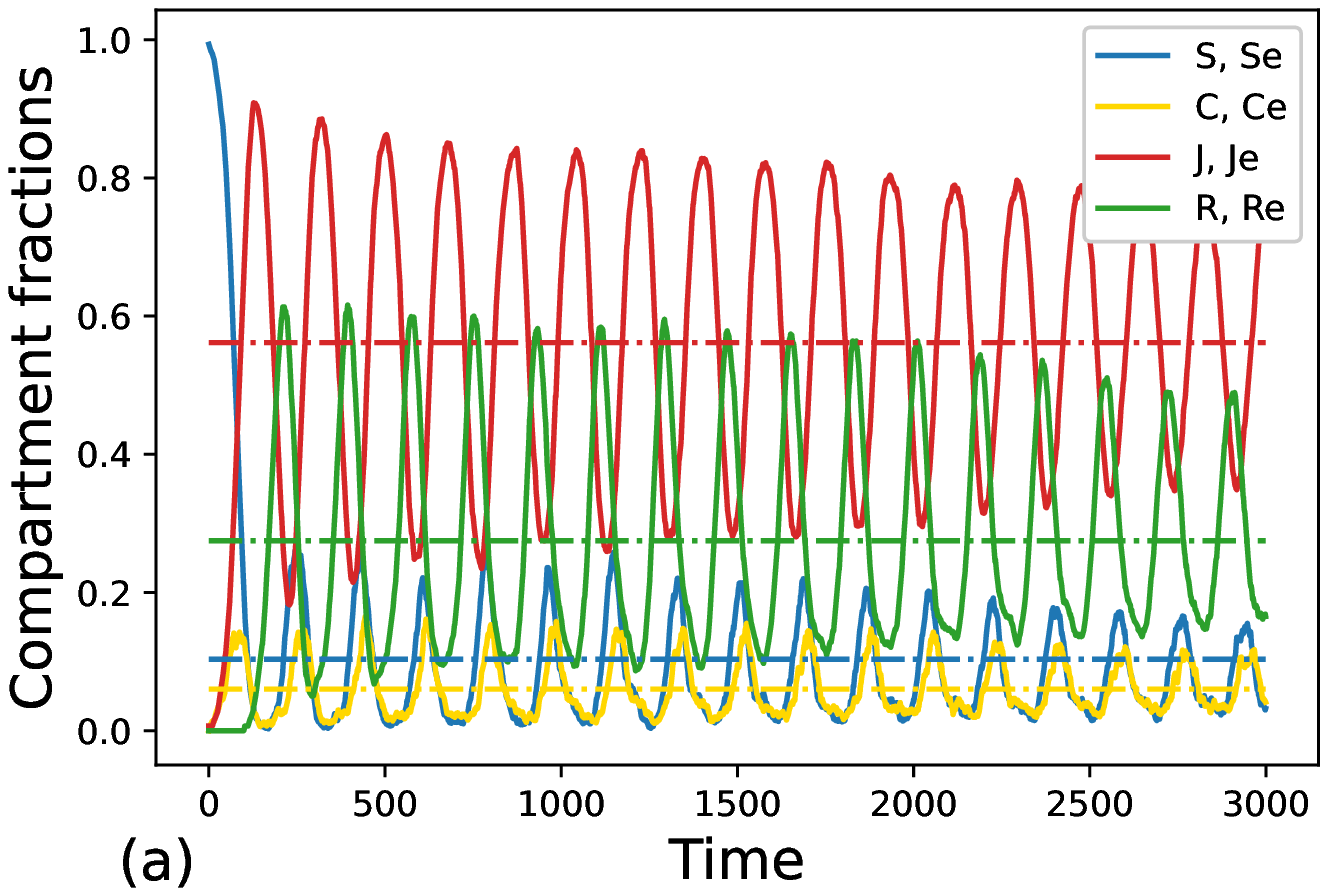}
\includegraphics[width=0.52\textwidth]{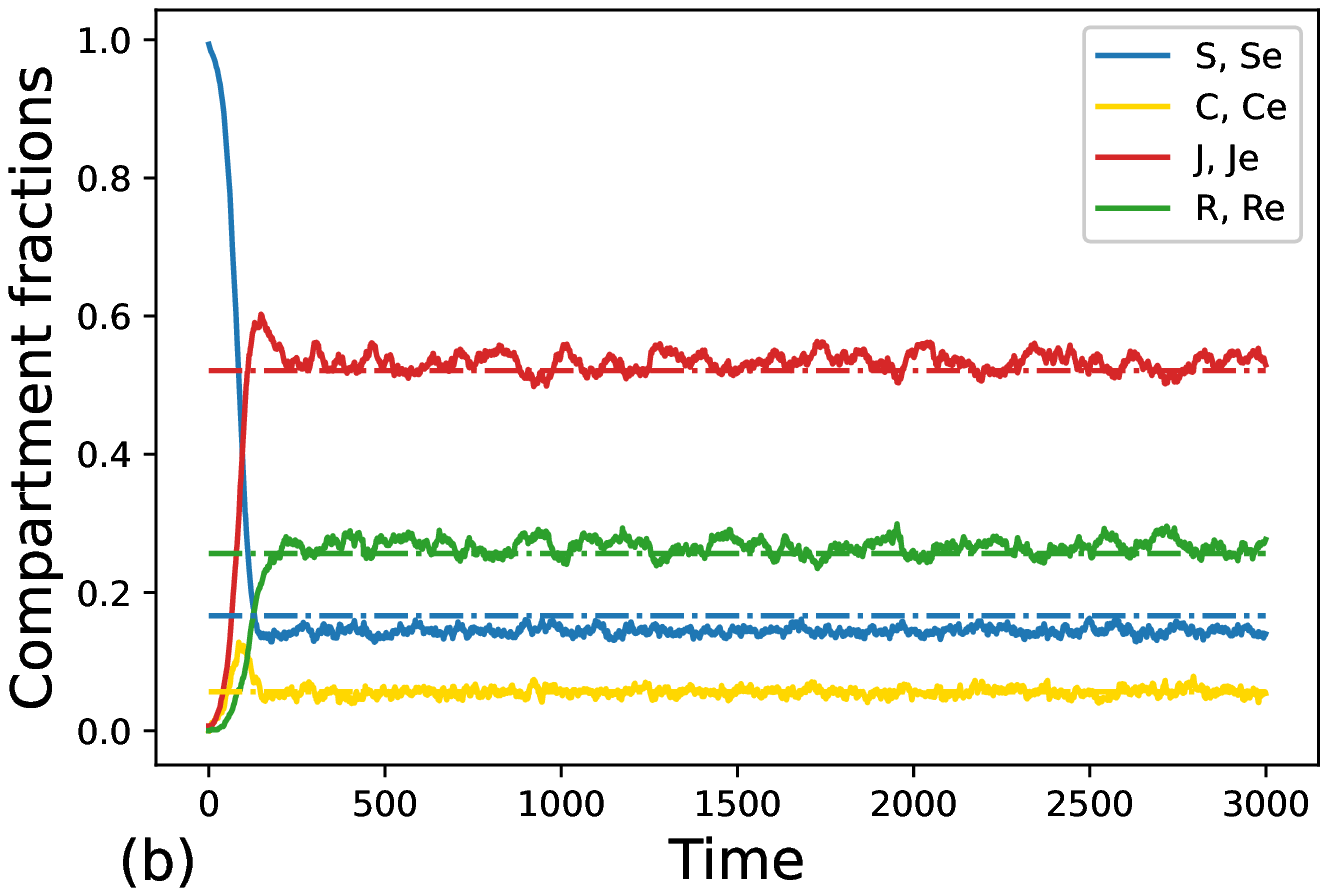}
\caption{\label{actuel} Compartment fractions averaged over 10 random walk realizations for (a) $\delta$-distributed 
and (b) exponentially distributed waiting times
with
$Z=150$, $N=21$ (density $Z/N^2\approx 0.34$), $P_{inf}=0.9$, 
and the means $\langle t_C\rangle=10$, $\langle t_I \rangle =100$, $\langle t_R\rangle =50 $.
\newline
Endemic states (dashed lines) for (a) $\delta$-distributed waiting times:
$S_e \approx 0.10$ ($R_0 \approx 9.68$), $C_e \approx 0.06$, $J_e \approx 0.56$, $R_e \approx 0.27$, and
$r_C \approx 0.07$, $r_I \approx 1.00$, $r_R=0.98$, see (\ref{model_ratios}).
\newline
Endemic states (dashed lines) for (b) exponential waiting times: $S_e\approx 0.16$ ($R_0\approx 6.01$), $C_e \approx 0.06$, $J_e\approx 0.52$, $R_e\approx 0.26 $ and $r_C\approx 1.07$, $r_I \approx 1.00$, $r_R\approx 0.98$
}
\end{figure}
\noindent The depicted time evolutions of Figures \ref{actuel} exhibit for
$\delta$-distributed waiting times slightly attenuated oscillations. This indicates that this case is close to an oscillatory (Hopf-) instability.
We approach this instability by slightly reducing the walkers density (reducing $R_0$). The corresponding persistent oscillatory behavior is shown in subsequent Figure \ref{actuel2}(a). 
\begin{figure}[H]
\includegraphics[width=0.52\textwidth]{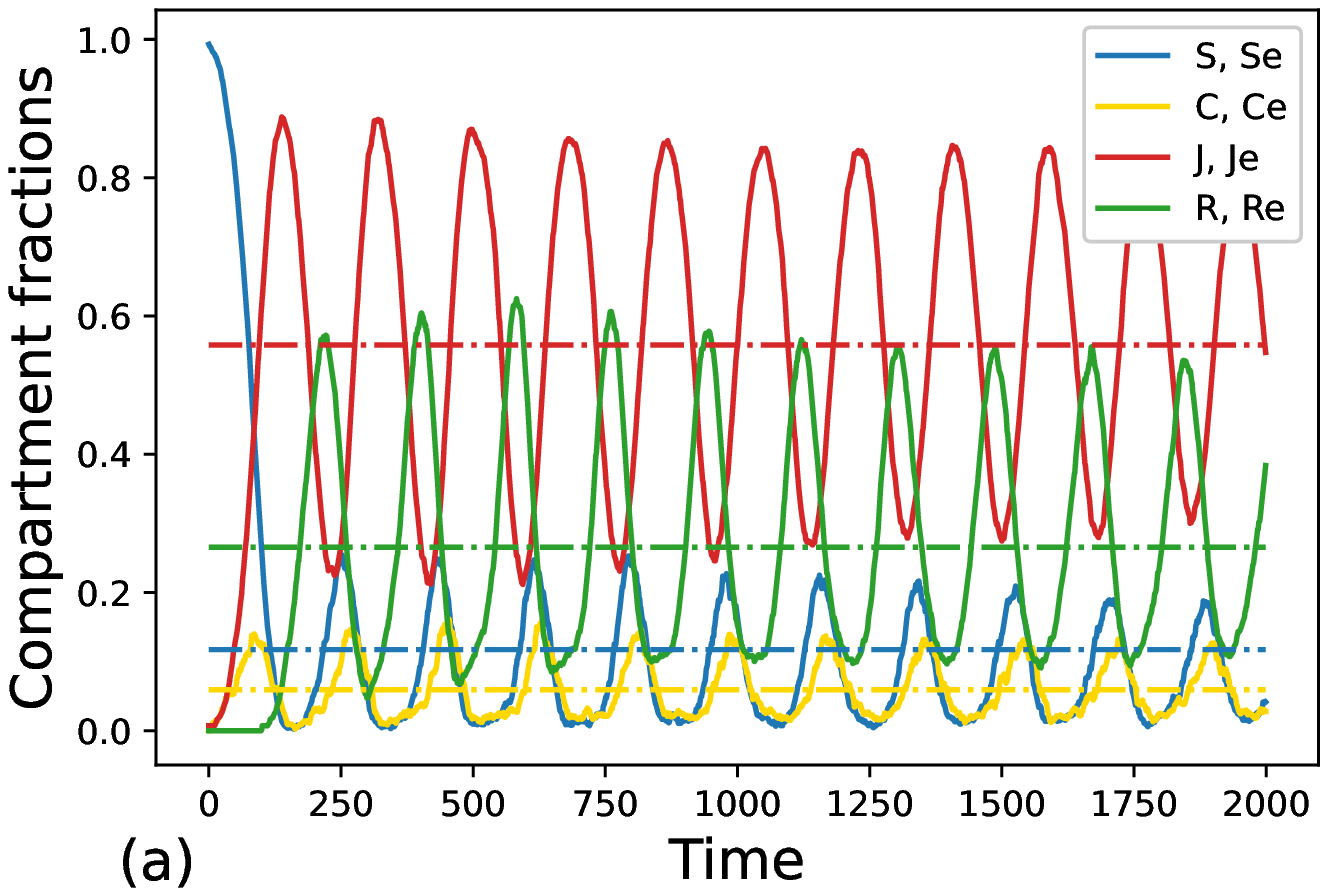}
\includegraphics[width=0.52\textwidth]{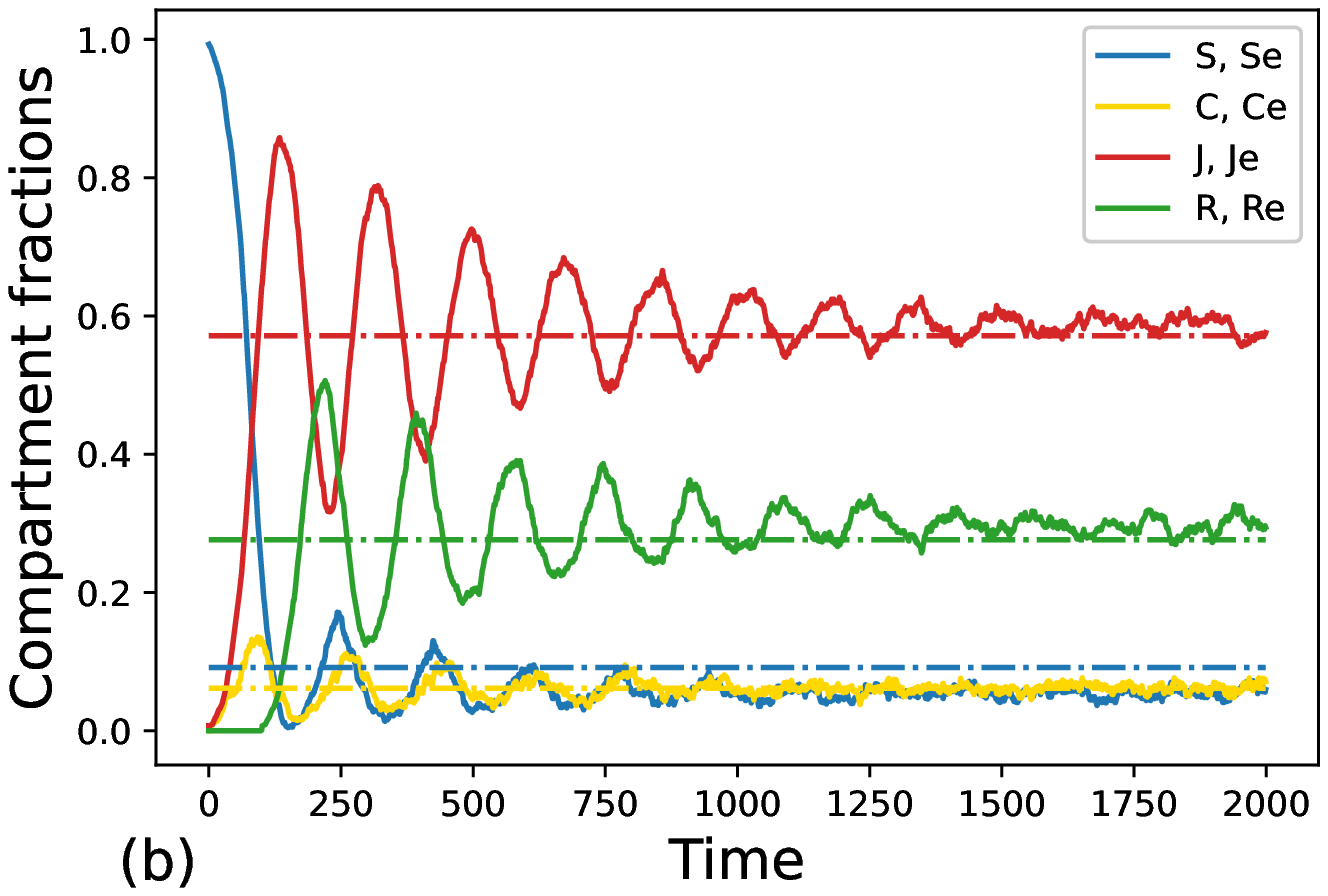}
\caption{\label{actuel2} (a) $\delta$-distributed waiting times, all parameters as in Figure \ref{actuel}(a) but with
$Z=140$ to have a slightly lower density to approach the oscillatory (Hopf) instability (density $Z/N^2 \approx 0.32$). Endemic values:
$S_e \approx 0.11$ $(R_0 \approx 8.53$), $C_e\approx 0.06$, $J_e \approx 0.56$, 
$R_e\approx 0.26$, and ratios
$r_C \approx 1.07$, $r_I \approx 1.01$, $r_R \approx 0.96$. \newline
(b) Identical parameters (including density) as in (a) but with exponentially distributed $t_C$ with $\langle t_C \rangle =10$, 
Gamma distributed $t_R$ with $\xi_R=0.1$ and $\langle t_R \rangle =50$ and $t_I$ is $\delta$-distributed ($t_I=100$). The different types of distributions decrease slightly $S_e\approx 0.09$ with increasing $R_0 \approx 10.95$ compared to (a) leading to attenuated oscillations. Other endemic values $C_e\approx 0.06$, $J_e \approx 0.57$, $R_e\approx 0.28$, and ratios
$r_C \approx 1.08 $, $r_I \approx 1.00 $, $r_R\approx 0.97$ } 
\end{figure}
\noindent Both plots in Figure \ref{actuel2} differ by changing to exponentially distributed $t_C,$ and Gamma distributed $t_R$. 
All other parameters including the density are identical in both Figures \ref{actuel2}(a,b) where \ref{actuel2}(a) exhibits an Hopf unstable persistent oscillatory behavior.
The resulting $S_e$ in Figure \ref{actuel2}(b) is slightly lower increasing $R_0$ and leading to attenuation of the Hopf oscillations. This demonstrates as both plots employ identical parameters but different waiting time PDFs that $R_0$ also depends on further characteristics such as for instance the variance.

\vspace{1cm}

\begin{figure}[H]
\centerline{\includegraphics[width=0.75\textwidth]{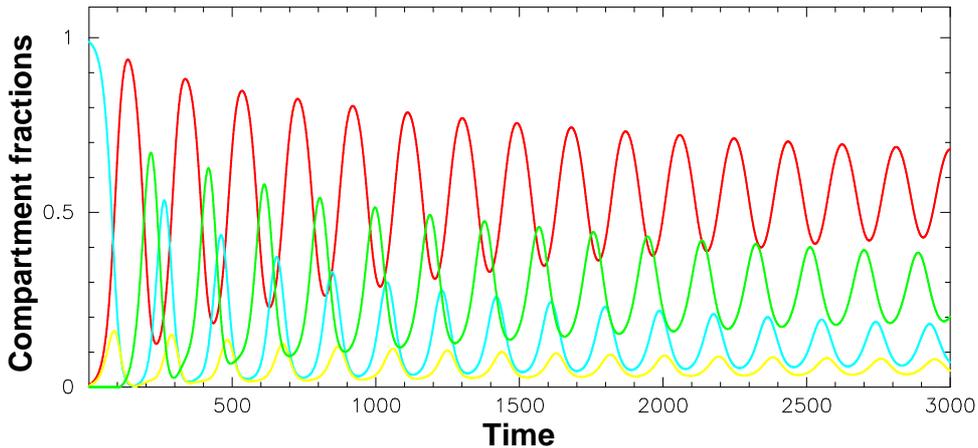}}
  \caption{\label{fig8-new}  Numerical solution of the ODE model (\ref{SRIRS-delta}) with
    the parameters of Figure \ref{actuel2}(a) and $R_0=9.1,\ j_0=0.01,\ s_0=0.99,
    \ c_0=r_0=0$. The color code is the same as in Figure \ref{actuel2}.}
\end{figure}

\noindent
The plot in Figure \ref{fig8-new} shows a numerical solution of the macroscopic system
(\ref{SRIRS-delta}) for the same parameters as in Figures \ref{actuel}(a) and \ref{actuel2}(a). The at least
qualitative similarity to the microscopic result is impressive (see especially Figure \ref{actuel2}(b) exhibiting attenuated oscillations). In both models the
number of infectious dominates due to the relatively long infection time. In the
long time limit the oscillations are damped and the endemic equilibrium is asymptotically reached.
\begin{figure}[H]
\centerline{\includegraphics[width=0.75\textwidth]{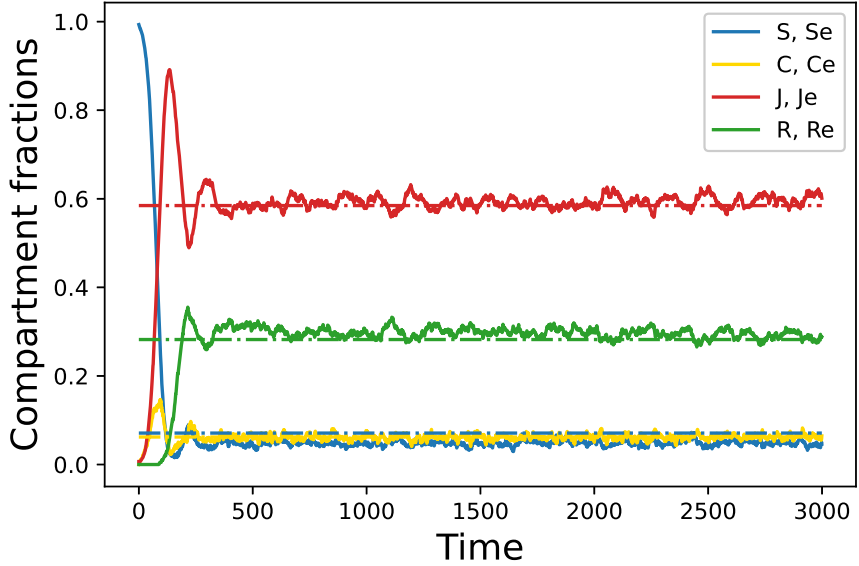}}
\caption{\label{Erlang} Compartment fraction averaged over 10 random walk realizations and 
Gamma distributed waiting times.
Same parameters as in Figure \ref{actuel} except $\xi_C=0.5$, $\xi_I=0.2$, $\xi_R=0.01$, \newline
Endemic state (dashed lines): $S_e \approx 0.07$ ($R_0 \approx 14.09$), $C_e \approx 0.06$, $J_e \approx 0.58$, $R_e \approx 0.28$ and ratios
$r_C \approx 1.06$, $r_I \approx 1.00$, $R_R \approx 0.97$.}
\end{figure}
\begin{figure}[H]
\includegraphics[width=0.52\textwidth]{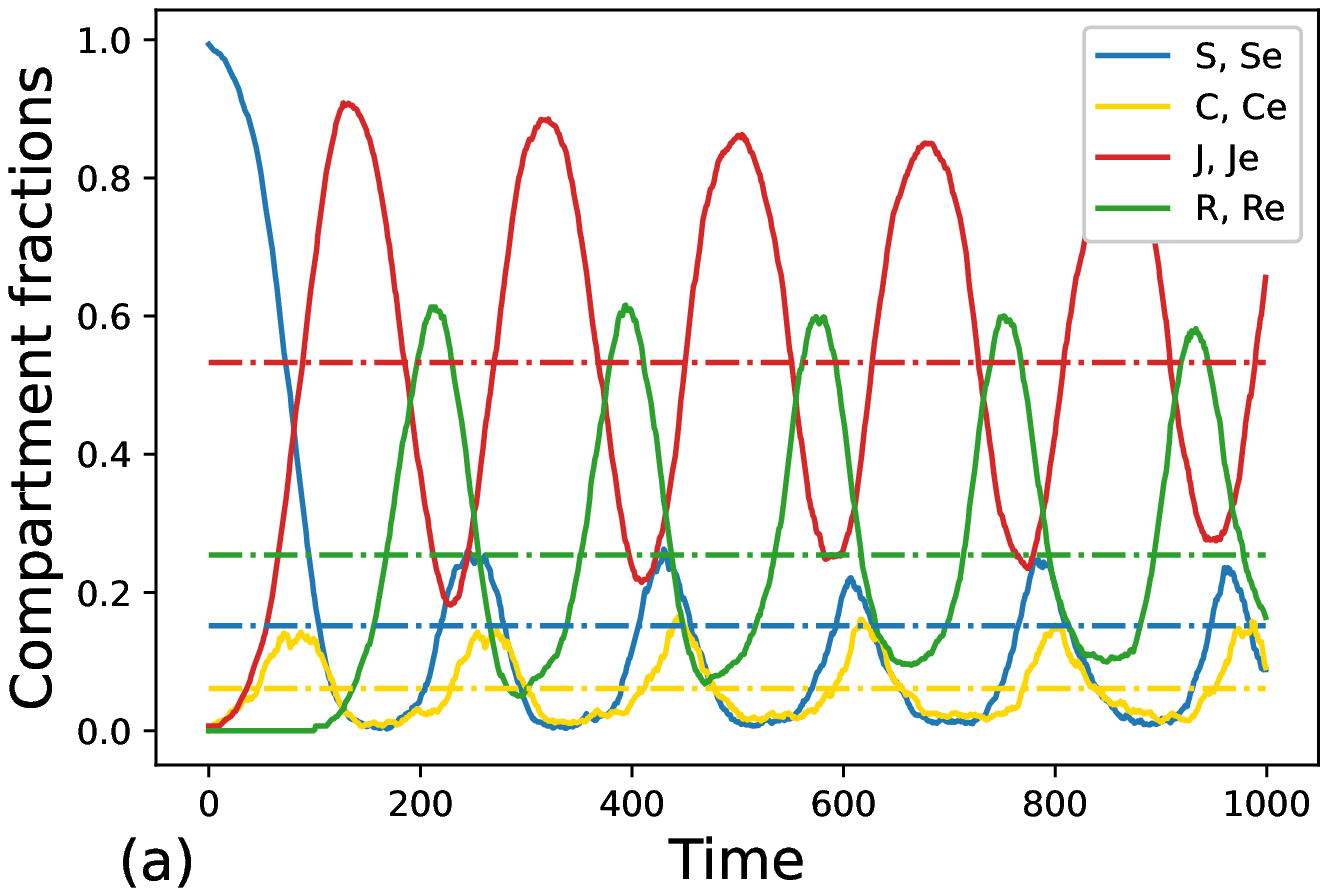}
\includegraphics[width=0.52\textwidth]{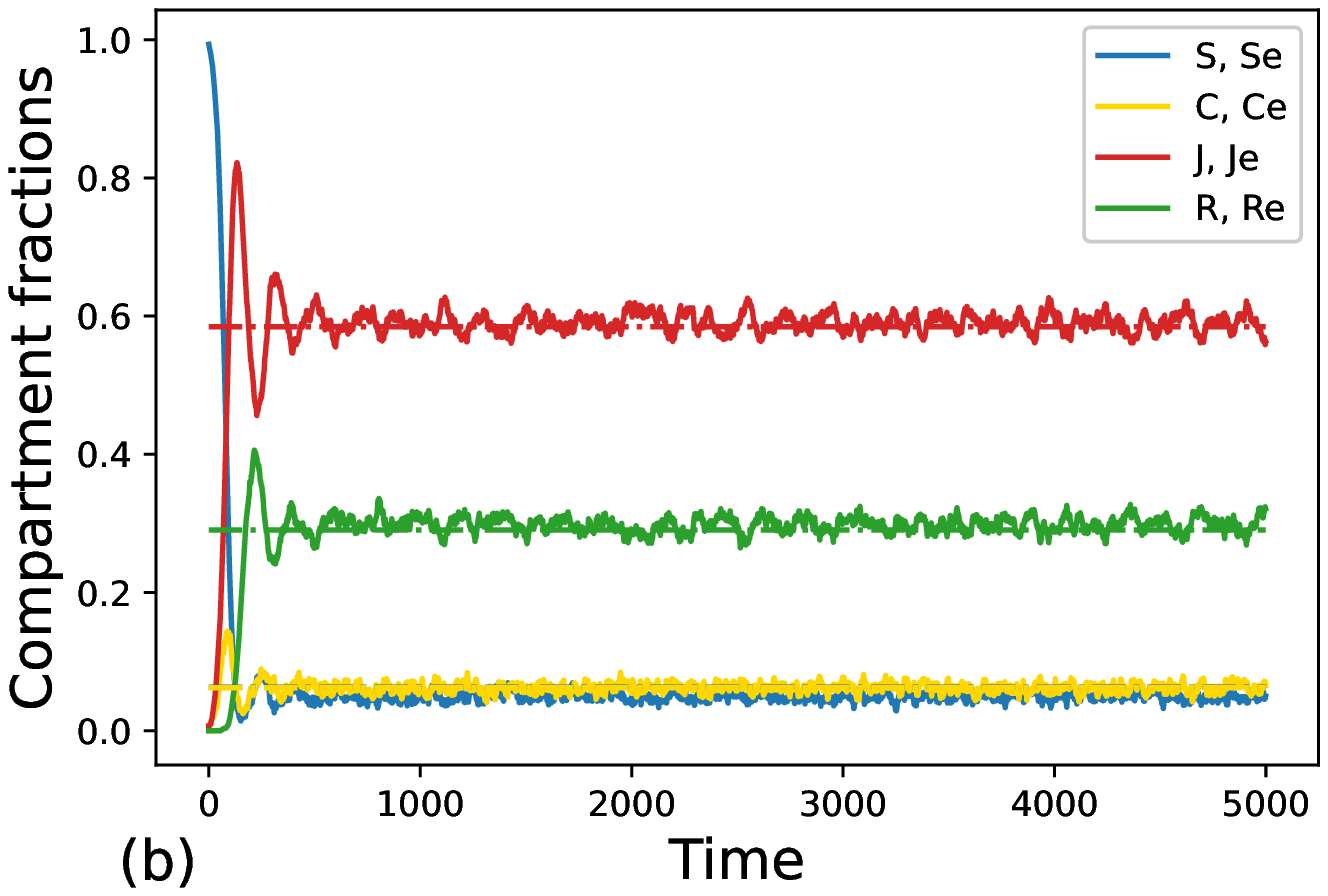}
\caption{\label{deltas}(a) $\delta$-distributed waiting times all parameters are identical as in (b), except $\xi_{C,I,R} = 10^4$ (generating the $\delta$-distribution, see (\ref{Gamma-dis-deltalimit})). Endemic values (dashed lines): $S_e\approx 0.15 \, (R_0=1/S_e\approx 6.59), C_e\approx 0.06, J_e\approx 0.53, R_e \approx 0.25$ with ratios
$r_C\approx 1.15$, $r_I\approx 1.00$, $r_R \approx 0.96$. \newline
(b): All waiting times are Gamma distributed where their means are identical with those of (a) with $\langle t_C \rangle = 10$, $\xi_C=0.1$, $\langle t_I \rangle = 100$, $\xi_I=0.1$ ($\alpha_I=10$), $\langle t_R \rangle = 50$, $\xi_R=0.05$, $Z=150$, $N=21$ (density $Z/N^2 \approx 0.34$), $P_{inf}=0.9$,
$S_e\approx 0.06 (R_0=1/S_e\approx 15.91), C_e\approx 0.06, J_e\approx 0.58, R_e \approx 0.29$, and ratios
$r_C\approx 1.06$, $r_I\approx 0.99 $, $r_R \approx 0.99$.}
\end{figure}
\noindent
In the simulations of Figures \ref{Erlang} and \ref{deltas} we consider various combinations of waiting time distributions. In Figure \ref{deltas} we compare
$\delta$-distributed waiting times and Gamma distributed waiting times with the same means. In Figure \ref{deltas}(a) $r_C$ has with $15\%$ a relative
large deviation from Eq. (\ref{endem}). This can be explained as 
as (\ref{S_e}) is an asymptotic relation converging for large $t$ to 
the endemic values where in Figure \ref{deltas}(a) the time range is not very large.
In Figure \ref{deltas}(b) the time range is considerably increased, thus
all ratios $r_{C,I,R}$ become very close to one, indicating excellent accordance with
Eq. (\ref{endem}).
The slightly different absolute values are due to slightly different $R_0=1/S_e$ despite all parameters $P_{inf}$, $Z$, $N$ are identical. This confirms our observation in Figures \ref{actuel2}
that $R_0$ depends on further characteristics of the waiting time PDFs such as for instance their variances.

\vspace{1cm}

\begin{figure}[H]
\centerline{\includegraphics[width=0.75\textwidth]{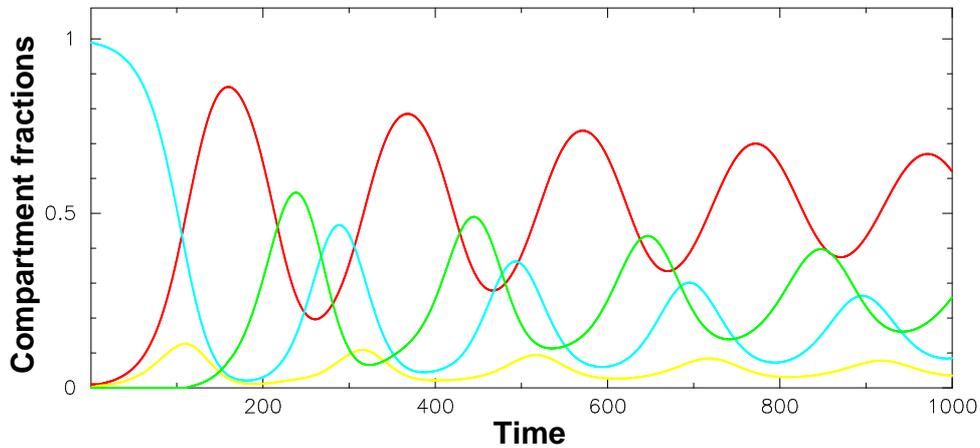}}
  \caption{\label{fig11}  Numerical solution of the ODE model (\ref{SRIRS-delta}) with
    the parameters of Figure \ref{deltas}(a) and $R_0=6.6,\ j_0=0.01,\ s_0=0.99,
    \ c_0=r_0=0$.}
\end{figure}
\noindent Figure \ref{fig11} shows again the numerical solution of the macroscopic system
(\ref{SRIRS-delta}) which are here for the smaller $R_0$ again in good agreement
with the averaged microscopic behavior.

\section{Conclusions}
We studied a macroscopic four compartment SCIRS model with memory effects introduced by random compartmental waiting times. 
We derived evolution equations for different arbitrary waiting time distributions and considered pertinent cases such as exponential (memoryless), Mittag Leffler (fat-tailed with long memory), and sharp ($\delta$-distributed) waiting times. We highlighted connections with general fractional calculus and showed that the evolution equations are of general (fractional) type (Eqs. (\ref{SCIRS-model_memory}), (\ref{SCIRS-model_memory_RL})). For waiting time PDFs with existing mean we obtained exact formulas for the endemic equilibrium and conditions of its existence and identified 
the `basic reproduction number' $R_0=\beta \langle t_I \rangle$ as crucial parameter 
controlling whether or not a SCIRS epidemic starts to spread.

We also found (Appendix \ref{R0_finite_means}) that the healthy state always is unstable if $t_I$ has a fat-tailed PDF regardless of the distributions of $t_{C,R}$. In this case $R_0$ and $\langle t_I\rangle$ do not exist.
We interpret this instability by the occurrence of very long infectious periods $t_I$ strongly boosting the epidemic spreading.
Due to their general importance, the time-fractional cases call for further thorough investigations.

We compared the macroscopic SCIRS model with a random walk approach. Our computer experiments have given numerical evidence that Eqs. (\ref{endem}) are fulfilled for any of the implemented waiting time distributions (with initial conditions very close to the healthy state) whenever the endemic equilibrium exists. This is further confirmed by comparison of results produced by the random walk simulations and by direct numerical integration of Eqs. (\ref{SRIRS-delta}) for two cases of $\delta$-distributed waiting times, see Figures \ref{actuel} (left), \ref{fig8-new} and \ref{deltas} (left), \ref{fig11}, respectively.

All these results give strong evidence that random walks offer an appropriate microscopic picture of the SCIRS dynamics.
It is also found in the simulations that waiting time PDFs with different variances
but otherwise identical parameters may influence $R_0$. 

For future research it would be desirable to relate the endemic value $S_e$ (i.e. $R_0$) 
(here determined numerically) with characteristics of the waiting time PDFs (such as their variances and others), infection probabilities in a collision of I and S walkers and of the random walk.

An interesting extension of our model could be considering the cases of a non-constant total population, i.e. when we admit on the one hand birth and death rates (i). On the other hand another promissing direction is to analyze how spatial inhomogeneities affect the epidemic dynamics (ii). 
In the first case (i) one can introduce a further compartment D (deaths) where a walker which is in state I
can either perform after a random waiting time $t_I$ a transition to R as in the present version of the model or a transition to D after a random waiting time $t_D$. The transition rate to D is the mortality rate due to the disease. The walkers having made a transition to R continue their transition pathway as in the present version of the model (i.e. to S and so forth). The choice whether a walker who is in compartment I performs a transition to D or R could be assumed random (with a certain probability) as well. Intuitively one might expect that the cases of persistent oscillatory behavior in the infected population of the present version of the model would turn into attenuated oscillations.
\\
Moreover, 
including spatial effects (ii) into the model leads to the occurrence of fluxes and inhomogeneities with macroscopic space-time partial differential evolution equations of general fractional types. 
The spatial effects involved are closely related to the assumption of admissible jumps of the walkers in their random walk, for instance local steps as in the present model or long-range steps as for instance in L\'evy flights \cite{MetzlerKlafter2000} or a combination of both defining different classes of random walkers (corresponding to distinct mobility of individuals). It would be desirable to study the connection of the random walk (beyond simple walks) and the macroscopic space-time evolution equations. The present SCIRS model would represent a pertinent benchmark limiting case in the mentioned extensions (i) and (ii).

A further challenge is the investigation of microscopic random walk models and their connection
with oscillatory (Hopf) instabilities (condition for persistent oscillatory behavior).
The characteristics of the Hopf instabilities of the healthy state and of quasi-periodic outbursts
are closely related with the specific distributions and the mutual ratios of the waiting times $t_{C,I,R}$. Models of the SCIRS class as in the present paper may help decision makers to take improved protection measures such as confinement, wearing of masks, or vaccination. The effect of vaccination can be directly studied in our SCIRS model by considering 
the stability of initial states with a non vanishing number of immune (vaccinated) individuals. We shall leave these issues for future research.

Generally SCIRS type models with memory as introduced in the present paper may open a wide research field to better understand the phenomenology of real world epidemic dynamics.
\section*{Acknowledgement}
T.G. gratefully acknowledges to have been hosted at the Institut Jean le Rond d'Alembert for the sake of the present study and development of the Python codes. We thank the referees for useful comments which helped us to improve the presentation.

\begin{appendix}
\section{General derivatives}
\label{general_fractional_derivatives}
We can alternatively represent Eqs. (\ref{SCIRS-model_memory}) in the Riemann-Liouville sense of general derivatives as
\begin{equation}
 \label{SCIRS-model_memory_RL}
 \begin{array}{clr}
\displaystyle  \frac{d}{dt}s(t) & = \ds  - {\cal A}(t)  
+   \frac{d}{dt}\int_0^t{\cal L}_R(t-\tau)r(\tau){\rm d}\tau & \\  \\
\displaystyle \frac{d}{dt}c(t) & = \ds  {\cal A}(t) - \frac{d}{dt}\int_0^t{\cal L}_C(t-\tau) c(\tau){\rm d}\tau \\ \\
\displaystyle  \frac{d}{dt}j(t) & =\ds  \frac{d}{dt}\int_0^t{\cal L}_C(t-\tau) c(\tau){\rm d}\tau  - \frac{d}{dt}\int_0^t{\cal L}_I(t-\tau)[j(\tau)-j_0]{\rm d}\tau \\ \\
 \displaystyle  \frac{d}{dt}r(t) & =\ds  \frac{d}{dt}\int_0^t{\cal L}_I(t-\tau)[j(\tau)-j_0]{\rm d}\tau - \frac{d}{dt}\int_0^t{\cal L}_R(t-\tau)r(\tau){\rm d}\tau
 \end{array}
\end{equation}
which also is obtained straight-forwardly from the Laplace transformed representation (\ref{SCIRS-model-laplace}) and employing definition (\ref{memory}). On the right-hand sides general derivatives come into play. 
The notion of `general' (fractional) derivative was introduced by Kochubei \cite{Kochubei2011} generalizing 
$\frac{d}{dt}y(t)$ as (in the Caputo sense) 
\beq
\label{Kochubei_derivative}
D^*_t \cdot y(t) = \int_0^t k(t-\tau)\frac{d}{d\tau}y(\tau){\rm d}\tau = 
\frac{d}{d t} \int_0^t k(t-\tau)y(\tau){\rm d}\tau - y_0k(t) = D_t\cdot y(t)-y_0k(t) 
\eeq
with $y_0= y(t)\big|_{t=0}$.
The part $D_t\cdot y(t) = \int_0^t k(t-\tau)y(\tau){\rm d}\tau$
is a general derivative in the Riemann-Liouville sense
for some admissible kernels $k(\tau)$ (see \cite{Kochubei2011} for an outline of this theory). 
The general derivative (\ref{Kochubei_derivative}) has the LT
\beq
\label{general_der_laplace}
\int_0^{\infty} e^{-\lambda t} D_t y(t) {\rm d}t = \lambda{\hat k}(\lambda){\hat y}(\lambda)-y_0{\hat k}(\lambda)
\eeq
The general derivatives contain the class of standard fractional derivatives of Caputo and Riemann-Liouville 
type and for a $k(t)=\delta(t)$ the standard first order derivative.
\section{Stability analysis for arbitrary waiting time densities}
\label{R0_finite_means}
In Section \ref{delta_kernels} we have derived 
that the healthy state is unstable if $R_0=\beta t_I>1$ for $\delta$-distributed kernels.
Here we consider arbitrary waiting time kernels first with finite means to generalize this result and at the end of this appendix we briefly look at the time fractional case.
To this end we take into account that the PDFs $K_{C,I,R}(t)$ 
can be seen as a superposition (average over $t_{C,I,R}$) of the $\delta(t-t_{C,I,R})$-kernel as follows
\beq
\label{kernel_superpos}
K_{C,I,R}(t) = \langle \delta(t-t_{C,I,R}) \rangle = \int_0^{\infty} \delta(t-\tau) K_{C,I,R}(\tau){\rm d}\tau=  (\delta \star K_{C,I,R})(t)
\eeq
and in this way we can average
\beq
\label{function_average}
\langle f(t_{C,I,R}) \rangle = \int_0^{\infty}K_{C,I,R}(\tau) f(\tau){\rm d}\tau
\eeq
for sufficiently good functions $f$ thus
\beq
\label{averaging_exp}
\langle e^{-\mu t_{C,I,R}} \rangle ={\hat K}_{C,I,R}(\mu)
\eeq
yielding the Laplace transform of the PDF. 
Therefore, averaging Eqs. 
(\ref{SRIRS-delta}) for $\delta$-kernels over the waiting times $t_{C,I,R}$ brings us back to the general SCIRS Eqs. (\ref{SCIRS-model}) which can be seen from
\beq
\label{average_convol}
\left\langle {\cal A}(t-t_C) \right\rangle = \int_0^t {\cal A}(t-\tau)K_c(\tau){\rm d}\tau = (K_C \star {\cal A})(t)
\eeq
where we used causality of ${\cal A}$ and $K_C$.
Hence we can generalize the case of $\delta$-kernels (\ref{cond_eqs}) to any kernels by accounting for
$$ \langle e^{-\mu(t_I+t_R + t_C)} \rangle =  {\hat K}_C(\mu){\hat K}_I(\mu){\hat K}_R(\mu) $$ where we always use the mutually independence of the waiting times.
Hence by averaging Eq. (\ref{cond_eqs}) we get
\beq
\label{gen_of_34}
\begin{array}{clr}
\ds  A_0(u,w)[1-{\hat K}_C(\mu){\hat K}_I(\mu){\hat K}_R(\mu))]+\mu u & =0 & \\ \\
\ds  A_0(u,w)[1-{\hat K}_C(\mu)] - \mu v & = 0 & \\ \\
\ds  A_0(u,w){\hat K}_C(\mu)[1-{\hat K}_I(\mu)] - \mu w & =0 &  \\ \\
\ds   A_0(u,w){\hat K}_C(\mu){\hat K}_I(\mu)[1-{\hat K}_R(\mu)] - x\mu  & = 0 &
\end{array}
\eeq
leading to the solvability condition
\beq
\label{det_solv_gen_PDF}
\left\| \begin{array}{cl}\ds  \beta J_e[1-{\hat K}_C(\mu){\hat K}_I(\mu){\hat K}_R(\mu)]+\mu ;& \ds \beta S_e[1-{\hat K}_C(\mu){\hat K}_I(\mu){\hat K}_R(\mu)] \\ \\   \ds \beta J_e {\hat K}_C(\mu)[1- {\hat K}_I(\mu)] ; & \ds \beta 
 S_e 
{\hat K}_C(\mu)[1-{\hat K}_I(\mu)] -\mu \end{array} \right\| \mu^2 = 0.
\eeq
We have again (as in (\ref{det_solv})) $\mu=0$ as a threefold eigenvalue. The remaining non-zero eigenvalue is determined by
\beq
\label{eigenval_gen}
\mu= \beta S_e {\hat K}_C(\mu)[1-{\hat K}_I(\mu)]- \beta J_e[1-{\hat K}_C(\mu){\hat K}_I(\mu){\hat K}_R(\mu)]
\eeq
We have to point out that in order to obtain these results we have relaxed 
causality in the exponential ansatz
(\ref{stability_analysis}) and assumed that it is defined for all $t\in \mathbb{R}$. The difference of causal and non-causal averaging can be seen by comparing
\beq
\label{average-exp}
\begin{array}{clr}
\ds \left\langle e^{\mu(t-t_C)} \right\rangle =  & \ds  e^{\mu t}\int_0^{\infty}e^{-\mu\tau} K_C(\tau){\rm d}\tau =  e^{\mu t} {\hat K}_C(\mu) & \\ \\ & \neq  \ds \left\langle \Theta(t-t_C) e^{\mu(t-t_C)} \right\rangle =  e^{\mu t}\int_0^{t}e^{-\mu\tau} K_C(\tau){\rm d}\tau &
\end{array}
\eeq
where $\Theta(\tau)$ indicates the Heaviside unit step function and corresponds to causality. To obtain the condition (\ref{det_solv_gen_PDF}) we have averaged 
as in the first line of (\ref{average-exp}) where the exponential is non-causal allowing exponentials with arguments $t-t_C<0$ to contribute. Both sides become asymptotically equal in the limit 
$t \to \infty$. Hence above relation (\ref{det_solv_gen_PDF}) can also be interpreted as the large time asymptotics of causal averaging.
Let us focus now on the stability of the healthy state $s_0=1$. Then (\ref{eigenval_gen}) 
yields ($S_e \to s_0=1$, $J_e \to j_0=0$) 
\beq
\label{eigenval_gen_healthy}
\mu = \beta {\hat K}_C(\mu)[1-{\hat K}_I(\mu)] = G(\mu).
\eeq
As in the case of $\delta$-distributed waiting times $G(\mu)$ is a concave function
with  $G(\mu) \to 0$ as $\mu \to \infty$ thus we can argue in the same way (see Eq. (\ref{stability_healthy}) 
and Figure \ref{Fig0}). Hence there is an intersection $\mu_0 > 0$ only
if $\frac{d}{d \mu}G(\mu)\big|_{\mu=0} >1$. Expanding $G(\mu)$ (assuming existing mean $\langle t_I\rangle$) gives
\beq
\label{right_handside_solv}
G(\mu) = \beta [1-\langle t_C \rangle \mu  +o(\mu)]  \langle t_I \rangle \mu
=  \beta \langle t_I \rangle \mu +o(\mu).
\eeq
Therefore,
\beq
\label{R0_factor}
\frac{d}{d \mu}G(\mu)\big|_{\mu=0} =: R_0 = \beta \langle t_I \rangle >1 
\eeq
is the condition for this instability of the healthy state and hence the interpretation as `basic reproduction number' $R_0$ makes sense. This holds for waiting time PDFs with existing means.
On the other hand relation (\ref{eigenval_gen_healthy}) holds for any waiting time PDF including fat-tailed ones which we briefly consider in the following. 
\newline\newline
\noindent{\it Time fractional case}
\newline\newline
Assuming $K_I(t)$ is a fat-tailed PDF such as Mittag-Leffler with
${\hat K}_I(\mu)=\frac{\xi_I}{\xi_I+\mu^{\nu}}$ , $\nu \in (0,1)$
and no matter whether or not the other kernels have existing means
we get
for $G(\mu)$ 
\beq
\label{t_fractional}
G(\mu) =\frac{\beta}{\xi_I}\, \mu^{\nu} + o(\mu^{\nu}) > \mu \hspace{2cm} (\mu \to 0)
\eeq
where $o(\mu^{\nu})$ indicates orders higher than $\nu$.
Thus (\ref{eigenval_gen_healthy}) (independent of $\frac{\beta}{\xi_I}$) always has a positive intersection since $\nu \in (0,1)$.
{\it We conclude that for fat-tailed $K_I(t)$ the healthy state always is unstable regardless of the distributions of $t_C$ and $t_R$}. Physically this instability can be understood by the occurrence of very long infectious times $t_I$ strongly boosting the epidemic spreading (corresponding to the limit $R_0, \langle t_I \rangle \to \infty$).

\end{appendix}

\end{document}